\documentclass[prb,twocolumn,showpacs,amsfonts]{revtex4}

\usepackage{epsfig}

\begin{document}

\title{Continuous Time Monte Carlo and Spatial Ordering in  Driven Lattice Gases: 
Application to Driven Vortices in Periodic Superconducting Networks}


\author{Violeta Gotcheva}
\altaffiliation{present address: Laboratory for Laser Energetics, University of Rochester,
Rochester, NY 14627, USA}
\affiliation{Department of Physics and Astronomy, University of
Rochester, Rochester, NY 14627, USA}
\author{Yanting Wang}
\altaffiliation{present address: Department of Chemistry, University of
Utah, Salt Lake City, UT  84112-0830, USA}
\author{Albert T.~J.~Wang}
\altaffiliation{present address: Department of Physics, MIT, Cambridge, MA 02139-4307, USA}
\author{S. Teitel}
\affiliation{Department of Physics and Astronomy, University of
Rochester, Rochester, NY 14627, USA}
\date{\today}

\begin{abstract}
We consider the two dimensional (2D) classical lattice Coulomb gas as a model
for magnetic field induced vortices in 2D superconducting networks.
Two different dynamical rules are introduced to investigate driven diffusive
steady states far from equilibrium as a function of temperature and driving force.  
The resulting steady states differ dramatically depending on which
dynamical rule is used. We show that the commonly used driven diffusive 
Metropolis Monte Carlo dynamics contains unphysical intrinsic randomness that destroys the
spatial ordering present in equilibrium (the vortex lattice) over most of the driven phase
diagram.  A continuous time Monte Carlo (CTMC) is then developed, which
results in spatially ordered driven states at low temperature in finite sized systems.  
We show that CTMC is the natural discretization of continuum Langevin dynamics, 
and argue that it gives the correct physical behavior when the discrete grid 
represents the minima of a periodic potential.
We use detailed finite size scaling methods to analyze the spatial structure of 
the steady states.  We find that finite size effects can be subtle and that very long
simulation times can be needed to arrive at the correct steady state.
For particles moving on a triangular grid, we find that the ordered
moving state is a transversely pinned smectic that becomes unstable to an anisotropic liquid on
sufficiently large length scales.  For particles moving on a square grid, the moving
state is a similar smectic at large drives, but we find evidence for a possible moving
solid at lower drives.  We find that the driven liquid on the square grid has
long range hexatic order, and we explain this as a specifically non-equilibrium effect.
We show that, in the liquid, fluctuations about the average center of mass motion
are diffusive in both the transverse and longitudinal directions.
\end{abstract}
\maketitle

\section{Introduction}

While the theory of phase transitions of systems in thermodynamic equilibrium
is a well established and mature area of statistical physics, much less is established
about analogous critical behavior in driven steady states far from equilibrium.
As has been the case in the study of equilibrium phase transitions, the use of lattice models,
in which the degrees of freedom are constrained to sit on the sites of a discrete periodic
grid, has led to analytical simplifications and greater accuracy in numerical simulations
for investigating such steady states, as compared to corresponding continuum models \cite{Schmittmann}.  One advantage of a lattice gas model for numerical 
simulations of driven interacting many-body systems is that particles hop in discrete jumps.  If a 
particle sits in a local potential minimum, the lattice gas dynamics can allow the
particle to hop over the energy barrier out of the minimum in a single move.
In contrast, in continuum simulations like molecular dynamics, considerable simulation
time can be wasted at low temperatures waiting for a thermal excitation that will
excite the particle over the energy barrier.  The lattice gas method can therefore
hope to simulate out to much longer effective times, and focus on effects due
to many-body interactions rather than single body potentials.

One of the first, and still one of the most commonly used, numerical methods
to simulate driven steady states of a lattice gas is the {\it driven diffusive Monte Carlo}
method.  This method, introduced by Katz, Leibowitz and Spohn \cite{DDMC}, 
extends familiar equilibrium Monte Carlo methods to the case of driven non-equilibrium
states.  The key idea of this method is to include the work done by the driving force 
on a moved particle, in addition to the change in interaction energy, when computing
the energy difference to use in the Monte Carlo test for making moves.
Such a term biases motion in the direction of the driving force, and, with the use
of periodic boundary conditions, results in a steady state with a finite particle current.
This algorithm, which satisfies local detailed balance for individual particle moves, 
seeks to model diffusively moving particles in the limit where motion is
dominated by thermal activation over energy barriers, rather than microscopic
dynamics.  The hope is that the main qualitative features of the driven steady states,
and possible phase transitions between them, will be independent of the details of
the microscopic dynamics, and so will be captured by this algorithm.

However, unlike equilibrium simulations, where {\it any} dynamics that satisfies
detailed balance is sufficient to generate the correct equilibrium
Gibbs ensemble and so equilibrium averages are in principal independent of the
microscopic dynamics, there is no such guarantee for non-equilibrium states.
Even for sets of dynamics that would appear to lie within the same dynamic
``universality class" \cite{HalperinHohenberg} 
from the viewpoint of symmetry and conserved quantities,
averages  in driven steady states far from equilibrium may conceivably be
qualitatively, not just quantitatively, different.

In this work we test this notion explicitly by considering two different versions
of driven diffusive Monte Carlo dynamics, both intended to model
the overdamped diffusive limit.  We consider first (i) driven diffusive {\it Metropolis}
Monte Carlo dynamics \cite{DDMC,Schmittmann} 
(DDMMC), where the standard Metropolis method is
used to select attempted excitations and decide whether or not to accept them.
We then consider (ii) driven diffusive {\it continuous time} Monte Carlo
dynamics (CTMC), where the continuous time Monte Carlo method \cite{Bortz,Newman}
is used to make a rejectionless dynamics.  We believe that our work is the
first application of continuous time Monte Carlo in the context of driven
diffusive problems.  We apply these methods to the problem of the driven
two dimensional (2D) classical one component lattice Coulomb gas, which 
serves as a model for logarithmically interacting, magnetic field induced, vortices in 
periodic 2D superconducting networks \cite{Franz}.
We consider both the cases of a triangular
and a square grid of sites.  This model is of interest because it allows one
to consider the effect of a uniform driving force on a system which has
spatially ordered states in equilibrium (the
vortex lattice), in contrast to simpler nearest neighboring Ising-like lattice
gas models \cite{Kwak}, which in general have no such periodic spatial order.  

We find that our two
dynamics result in dramatically different driven steady states,
when the system is acted upon by a uniform applied force $F$.
We find that over most of the $T-F$ phase diagram, the DDMMC method
results in a spatially disordered moving steady state with a very
short translational correlation length.  We argue that this behavior is due
to intrinsic randomness in the DDMMC algorithm, that is sufficient to 
disorder the moving system even at $T=0$.  In contrast, we find that the
CTMC method, at least for finite size systems, can result in spatially ordered
moving steady states, as well as orientationally ordered moving liquids.
We demonstrate that the CTMC method is the correct discretization of 
diffusive Langevin dynamics in a certain limit, and argue that it more generally 
describes motion when the discrete grid is thought of as representing the
minima of a one body periodic potential, and the energy barriers of this
potential are large compared to the energy change of hopping between minima.
Thus we believe that CTMC is not only a more interesting dynamics, but
also a more physically correct one.  For CTMC dynamics, we carry out detailed
finite size scaling analyses of our ordered steady states, and show that 
there can be subtle finite size effects due to diverging correlation lengths at
low temperatures.  We also show that exceedingly long simulations are needed,
in some cases, in order to arrive at the correct steady state distribution. 
 
The remainder of this paper is organized as follows.  In section \ref{sModels}
we define in detail our Coloumb gas model and our two lattice gas dynamics.
We discuss qualitative behaviors to be expected at low temperatures,
and define the observables we will measure.  In section \ref{sResultsTri}
we present the results of our simulations on a triangular grid of sites.
We show the phase diagrams of both the DDMMC and CTMC for a system of a
given finite size, and demonstrate the dramatic difference between them.  
We then focus the remainder of our work on CTMC.
We carry out detailed finite size scaling analyses to study the
structural order of the moving steady states in both the high drive and low drive
limits.  At low temperature we find an ordered moving smectic state, however
we argue that this state is ultimately unstable to a liquid on sufficiently large length
scales.  We also present results for dynamic behavior, studying the average
velocity of the system and the diffusion of the system about its center of mass motion.
In section \ref{sSquareResults} we present our results for simulations on a square
grid of sites.  We study several specific points in the phase diagram representative
of the high drive and low drive limits.  Unlike for the triangular grid, we find that
the structure of the ordered moving state appears to have different periodicities 
at different driving
forces.  We carry out finite size scaling to investigate the  stability of the ordered
states in the large size limit.  We show that, unlike the liquid state in equilibrium,
the liquid driven steady state possesses long range hexatic orientational order.
In section \ref{sConc} we discuss our results and present our conclusions.

Some aspects of this work, focusing on the differences between
DDMMC and CTMC and the structural order of driven states on the triangular
grid, have previously appeared as a letter \cite{Gotcheva}.  The detailed discussion
of the phase diagram on the triangular grid, the finite size scaling analyses,
the discussion of dynamical behavior, and all results for the square grid, are 
presented for the first time in the current work.

\section{Model and Methods}
\label{sModels}
\subsection{Coulomb Gas Model}

Our model is a classical one component lattice Coulomb gas of 2D interacting charges,
which may be taken as a model for interacting vortices in a 2D
superconducting network \cite{Franz}.  The charges are constrained to sit on the 
discrete sites $i$ of a periodic 2D grid.  If the basis vectors of the grid are
$\{\hat a_1,\hat a_2\}$, we take the grid to have finite length $L_\mu$ in
direction $\hat a_\mu$ and we take periodic boundary conditions in both directions.
The Hamiltonian is given by \cite{Franz},
\begin{equation}
{\cal H}={1\over 2}\sum_{i,j}(n_i-f)G({\bf r}_i-{\bf r}_j)(n_j-f)\enspace,
\label{eH}
\end{equation}
where the sum is over all pairs of sites $i$, $j$ of the grid, $n_i\in\{0,1\}$
is the integer charge on site $i$, $-f$ is a uniform neutralizing background charge,
and $G({\bf r})$ is the 2D lattice Coulomb potential which solves,
\begin{equation}
\Delta^2G({\bf r})=-2\pi\delta_{{\bf r},0}\enspace.
\label{eLL}
\end{equation}
where $\Delta^2$ is the discrete Laplacian for the grid.
Defining $\Delta^2$ appropriate to periodic boundary conditions
results in a $G({\bf r})$ that satisfies periodic boundary conditions.
For separations large compared to the grid spacing, but small 
compared to the grid length ($1\ll|{\bf r}|\ll L$), one has $G({\bf r})\simeq -\ln|{\bf r}|$.
The condition that the total energy remain finite imposes the charge
neutrality condition,
\begin{equation}
\sum_i n_i=fL_1L_2\equiv N_c\enspace.
\label{eneutrality}
\end{equation}

\begin{figure}
\epsfxsize=8.6truecm
\epsfbox{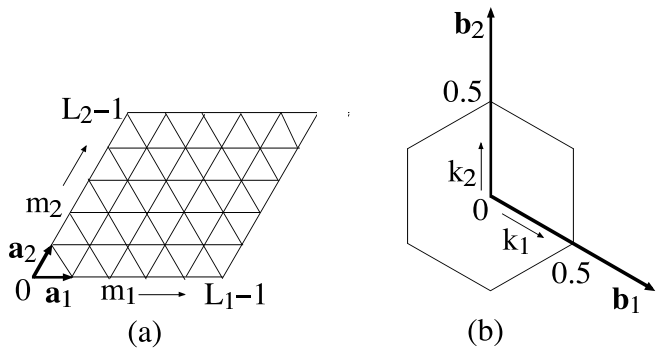}
\caption{(a) Triangular grid of size $L_1\times L_2$ with basis vectors $\hat a_1$ and $\hat a_2$.
(b) Reciprocal space to the triangular grid, with basis vectors ${\bf b}_1$
and ${\bf b}_2$.  Allowed wavevectors for Fourier transforms of real
space quantities can be restricted to the hexagonal first Brillouin zone shown in (b).
}
\label{f1}
\end{figure}

We will consider first the case of a triangular grid of sites.  Here
the basis vectors are,
\begin{eqnarray}
\hat a_1&=&\hat x\nonumber\\
\hat a_2&=&{1\over 2}\hat x+{\sqrt{3}\over 2}\hat y\enspace,
\label{eatri}
\end{eqnarray}
and the sites $i$ of the grid are specified by the position vectors,
\begin{eqnarray}
{\bf r}_i&=&m_1\hat a_1+m_2\hat a_2\enspace,\nonumber\\
m_\mu&\in&\{0,1,\dots,L_\mu-1\}\enspace.
\label{eri}
\end{eqnarray}
The geometry of this real space grid is illustrated in Fig.\,\ref{f1}a.

The solution to Eq.\,(\ref{eLL}) will be given in terms of its Fourier
transform.  The basis vectors of the reciprocal to the grid in Fourier transform
space are then,
\begin{eqnarray}
{\bf b}_1&=&2\pi\hat x-{2\pi\over\sqrt{3}}\hat y\nonumber\\
{\bf b}_2&=&{4\pi\over\sqrt{3}}\hat y\enspace,
\label{ebtri}
\end{eqnarray}
and the allowed wavevectors consistent with periodic boundary conditions are
given by,
\begin{eqnarray}
{\bf k}&=&k_1{\bf b}_1+k_2{\bf b}_2,\nonumber\\
k_\mu&\in&\{0,{1\over L_\mu},\dots,{L_\mu-1\over L_\mu}\}\enspace.
\label{ek}
\end{eqnarray}
Equivalently, one could translate the above wavevectors by an appropriate
linear combination of the basis vectors $\{{\bf b}_1,{\bf b}_2\}$ so that they
all lie in the hexagonal shaped first Brillouin zone of the reciprocal grid.  
The geometry of these allowed wavevectors is illustrated in Fig.\,\ref{f1}b.

Defining $\hat a_3\equiv \hat a_1-\hat a_2$, the discrete Laplacian for the
triangular grid is given by,
\begin{equation}
\Delta^2G({\bf r}) \equiv c\sum_{\mu=1}^3 \left[ G({\bf r}+\hat a_\mu)-2G({\bf r}) +G({\bf r}-\hat a_\mu\right)]
\label{eDLG}
\end{equation}
with $c$ an appropriate geometrical constant to give the correct continuum limit.
Taking the Fourier transform of the above, we find that the solution to Eq.\,(\ref{eLL}) is 
given by \cite{Franz,JRLee},
\begin{equation}
G({\bf r})={\pi\over cN}\sum_{{\bf k}\ne 0}{{\rm e}^{i{\bf k}\cdot{\bf r}}\over
3-\cos({\bf k}\cdot \hat a_1)-\cos({\bf k}\cdot \hat a_2)-\cos({\bf k}\cdot \hat a_3)}\enspace,
\label{eG}
\end{equation}
where $N=L_1L_2$ is the number of sites in the grid, and the sum is over all
the allowed wavevectors of Eq.\,(\ref{ek}).
The correct value of the geometric constant is $c=1/\sqrt{3}$.  However, in order to compare with
previous work done on this model \cite{Franz}, we will make the choice $c=2/3$.  This difference
amounts only to a rescaling of the temperature.

We will also discuss the case of a square grid of sites.  Here the real space
basis vectors of the grid are $\{\hat a_1,\hat a_2\}=\{\hat x,\hat y\}$,  the basis
vectors of the reciprocal space are $\{{\bf b}_1,{\bf b}_2\}=\{2\pi\hat x, 2\pi\hat y\}$,
and the lattice Coulomb potential is given by \cite{Franz,JRLee},
\begin{equation}
G({\bf r}) = {2\pi\over N}\sum_{{\bf k}\ne 0}{{\rm e}^{i{\bf k}\cdot{\bf r}}
\over 2-\cos({\bf k}\cdot\hat a_1)-\cos({\bf k}\cdot\hat a_2)}\enspace.
\label{eGs}
\end{equation}
\subsection{Lattice Gas Dynamics}

Our goal is to simulate the non-equilibrium steady states of the
lattice Coulomb gas when driven by a uniform applied force ${\bf F}$.
For {\it equilibrium} simulations, any dynamical rule that satisfies
detailed balance will converge to the correct equilibrium ensemble; 
the details of the dynamics may effect the speed of convergence,
but they are otherwise irrelevant for computing time independent thermodynamic averages.
For {\it non-equilibrium} steady states, however, even time independent averages 
may depend on the details of the microscopic dynamics.
Here we consider two different microscopic dynamics for the simple
case of over damped diffusively moving particles (the simplest case in the classification
scheme of dynamic critical phenomena by Halperin and Hohenberg \cite{HalperinHohenberg}).
Both dynamics involve single particle moves only.
We find that, for finite size systems, the resulting steady states for these
two dynamics can be qualitatively different.

\subsubsection{Driven Diffusive Metropolis Monte Carlo (DDMMC)}

The first lattice gas dynamics we consider is the commonly used
driven diffusive Metropolis Monte Carlo dynamics \cite{DDMC,Schmittmann} 
(DDMMC).  This algorithm
was introduced as a simple modification of ordinary equilibrium
Metropolis Monte Carlo.  It was intended to model the steady states of a 
driven system in the limit where motion is dominated by thermal
activation over energy barriers, and so presumably is not very sensitive 
to microscopic details.  The DDMMC algorithm is defined as follows.
At each step of the simulation a charge $n_i=1$ is selected at random, and the
charge is then moved a
test displacement $\Delta{\bf r}$ to a randomly chosen nearest
neighbor site.  For the triangular grid, $\Delta{\bf r}$ is chosen with
equal probability from the six possible directions $\pm \hat a_\mu$, 
$\mu=1,2,3$.  If ${\cal H}_{\rm old}$ and ${\cal H}_{\rm new}$ are
the interaction energies, Eq.\,(\ref{eH}), of the system before and after
this test move is made, one computes the energy difference,
\begin{equation}
\Delta U = {\cal H}_{\rm new}-{\cal H}_{\rm old} -{\bf F}\cdot\Delta{\bf r}
\enspace,
\label{eDU1}
\end{equation}
where the last term is the work done by the applied force on the moved charge.
This test move is then accepted or rejected according to the usual
Metropolis criterion,
\begin{equation}
\begin{array}{rl}
     {\rm accept\> if} &   \quad{\rm e}^{-\Delta U/T} >r \\
     {\rm reject\> if} &   \quad{\rm otherwise}\enspace,
\end{array}
\label{eMetro}
\end{equation}
where $r$ is a random variable uniformly distributed on $[0,1)$.
One pass of $N_c$ such steps equals one unit of simulation time.
The term in Eq.\,(\ref{eDU1}) proportional to the force ${\bf F}$
biases moves parallel to ${\bf F}$ and, in conjunction
with the periodic boundary conditions, 
will in general set up a steady state with a finite current of particles moving
parallel to ${\bf F}$.  Time independent averages are computed
in the usual Monte Carlo way, as a direct average over configurations
sampled every $N_{\rm pass}$ passes.

\subsubsection{Driven Diffusive Continuous Time Monte Carlo (CTMC)}
\label{sCTMC}

The second dynamics we consider, and the one which is used for the main part
of our work presented here, we call driven diffusive continuous time
Monte Carlo (CTMC).  The algorithm is defined as follows.  Starting from a
particular initial configuration, we denote 
by $(i\alpha)$ the single particle move of a  charge $n_i=1$ on site ${\bf r}_i$, to its
nearest neighbor site in direction $\hat \alpha$.  For the triangular grid,
$\hat \alpha\in \{\pm \hat a_1, \pm\hat a_2, \pm\hat a_3\}$.  For a grid in
which each site has $z$ nearest neighbors, the total number of such possible single particle
moves is $zN_c$.  For each such move, we compute the energy change $\Delta U_{i\alpha}$
according to Eq.\,(\ref{eDU1}), and define a probability rate for making this move,
\begin{equation}
W_{i\alpha}=W_0{\rm e}^{-\Delta U_{i\alpha}/2T}\enspace,
\label{eWialpha}
\end{equation}
where $1/W_0$ sets the unit of time.
Note that the above rates, as well as the Metropolis rates of DDMMC set by
Eq.\,(\ref{eMetro}), obey a local detailed balance.  If $s$ is an initial state,
and $s^\prime$ is the state reached from $s$ by making the single particle move $(i\alpha)$,
then,
\begin{equation}
{W(s\to s^\prime)\over W(s^\prime \to s)}= {\rm e}^{-\Delta U_{i\alpha}/T}\enspace.
\label{eDB}
\end{equation}
Although systems out of equilibrium do not in general need to satisfy
detailed balance, local detailed balance is physically reasonable if we
wish to regard each charge as moving in a local potential determined by its
interactions with the other charges and with the applied force.

Having specified the rates of Eq.\,(\ref{eWialpha}), we determine which move to 
make by regarding all $zN_c$ of the possible single particle moves as independent
Poisson processes. The probability that the next move will be $(i\alpha)$ is then,
\begin{equation}
P_{i\alpha}={W_{i\alpha}\over \sum_{(j\beta)}W_{j\beta}}\enspace,
\label{eP}
\end{equation}
and the average time it takes to make this move is,
\begin{equation}
\Delta t = {1\over W_{\rm tot}} \equiv {1\over \sum_{(j\beta)}W_{j\beta}} \enspace.
\label{eDt}
\end{equation}
We thus make a move by sampling the probability distribution $P_{i\alpha}$ of Eq.\,(\ref{eP}),
and then update the simulation clock by the amount $\Delta t$ of Eq.\,(\ref{eDt}).
Averages of observables ${\cal O}$ are computed as,
\begin{equation}
\langle{\cal O}\rangle = {1\over \tau}\int {\cal O}(t)dt = {1\over\tau}
\sum_s {\cal O}_s\Delta t_s\enspace,
\label{eO}
\end{equation}
where $s$ labels the steps of the simulation, ${\cal O}_s$ is the value of ${\cal O}$
in the configuration at step $s$, $\Delta t_s\equiv t_{s+1}-t_s$ is the time spent in the
configuration at step $s$ according to the simulation clock, and $\tau = \sum_s\Delta t_s$ is the 
total time of the simulation.  As in DDMMC, we will refer to $N_c$ simulation steps as one 
{\it pass} through the system.

The above algorithm is a straightforward extension of equilibrium
continuous time Monte Carlo \cite{Newman}, but we believe that this is its first application
in the context of driven non-equilibrium steady states.  The method was first
introduced as the ``n-fold way" for spin models \cite{Bortz}.  It owes its name to
the continuous variations in the time steps $\Delta t_s$, which vary from 
configuration to configuration, according to the energy barriers present in each
configuration.  It is a rejectionless algorithm designed to speed up equilibration
at low temperatures.  Rather than waste many rejected moves until a rare acceptance
takes one up and over an energy barrier, the energy barriers $\Delta U_{i\alpha}$
themselves set the time scale for each move, which then happens in a single
simulation step.  Simulation clock times can vary over orders of magnitude as
either $T$ or the height of the energy barriers vary.

In CTMC there are many possible choices for the rates $W_{i\alpha}$ that 
would satisfy local detailed balance.  In Appendix A we show that the particular
choice of Eq.\,(\ref{eWialpha}), with $W_0=cDT$ ($D$ is the diffusion constant), 
is the natural discretization to a periodic grid of sites of over damped continuum 
Langevin dynamics, and that the continuum limit is
reached when $\Delta U_{i\alpha}\ll T$.  Our simulations, however, being generally at
low $T$ or large ${\bf F}$, are mostly in the opposite limit of $\Delta U_{i\alpha}
\gtrsim T$.  To see what physical situation this limit corresponds to, consider a single
particle moving on a one dimensional grid of sites, in a driving force $F$.  
According to the CTMC algorithm, the average distance traveled in one step is,
\begin{equation}
\langle\Delta x\rangle = {{\rm e}^{F/2T}-{\rm e}^{-F/2T}\over {\rm e}^{F/2T}+{\rm e}^{-F/2T}}\enspace,
\label{eDx}
\end{equation}
while the average time for this step is given by,
\begin{equation}
{1\over \Delta t}= W_{\rm tot}=W_0\left[ {\rm e}^{F/2T}+{\rm e}^{-F/2T}\right]
\enspace,
\label{eDt2}
\end{equation}
leading to an average velocity,
\begin{equation}
\langle v\rangle={\langle\Delta x\rangle\over\Delta t}=2W_0\sinh(F/2T)\enspace.
\label{evel}
\end{equation}
At low ratios of $F/T$ the velocity is linear in the applied force,
$\langle v\rangle \simeq W_0F/T$; at large $F/T$, the velocity grows
exponentially, $\langle v\rangle \simeq W_0 {\rm e}^{F/2T}$.
We can compare the above result to that of an over damped particle moving in a continuum 
``washboard potential", $U(x)=-U_0\cos(2\pi x)-Fx$, which has been studied by
Ambegaokar and Halperin in the context of a driven Josephson junction \cite{Ambegaokar}.
The average velocity that they find
agrees exactly with Eq.\,(\ref{evel}) above, if one is in the limit
$T\ll 2 U_0$ and $F< 2\pi U_0$, and one identifies 
\begin{equation}
W_0=2\pi U_0 D \sqrt{1-\gamma^2}\,{\rm e}^{-2U_0\left[
\sqrt{1-\gamma^2}+\gamma \sin^{-1}\gamma\right]/T}\enspace,
\label{eWAH}
\end{equation}
where $D$ is the diffusion constant and $\gamma\equiv F/2\pi U_0$.
For small $\gamma$ the above $W_0$ reduces to a form independent of
the drive $F$,
\begin{equation}
W_0\simeq 2\pi U_0 D {\rm e}^{-2U_0/T}, \quad {\rm when\>}F\ll 2\pi U_0\enspace,
\label{eWAHs}
\end{equation}
which is the rate for activation over an energy barrier of $2U_0$.
CTMC thus describes the limit where the grid sites represent the minima
of a periodic pinning potential, and the applied force is weak enough that motion
is due to thermal activation of particles, one at a time, over the barriers of this periodic 
potential.  It is unclear  \cite{note} if CTMC can qualitatively describe the very large drive limit, 
$F\gg U_0$, where the washboard potential loses its local minima parallel to ${\bf F}$, and
the average velocity again becomes proportional to $F$.  
For the results reported in the following sections we will measure time
in units where $W_0=1$, independent of the temperature $T$ or driving force $F$.

\subsubsection{Behavior at Low Temperature}
\label{sLowT}

To get a better feel for the behavior of the above two lattice gas dynamic algorithms,
we can consider their behavior at low temperature.  In the limit $T\to 0$, the
DDMMC will reject all moves except those that lower the energy, i.e. $\Delta U_{i\alpha}<0$.
If one starts initially in the $F=0$ ground state and increases $F$, all moves will
be rejected until $F$ reaches a critical value $F_c$ equal to the interaction energy
associated with moving one charge forward parallel to ${\bf F}$.  The ordered ground state
charge lattice will therefore be pinned with $\langle {\bf v}\rangle =0$  for $F<F_c$, and 
moving with $\langle{\bf v}\rangle $ finite for $F>F_c$.  

Next we consider DDMMC at $T=0$ with $F\gg F_c$, so that the work done
by the force in Eq.\,(\ref{eDU1}) dominates the interaction energy $\Delta{\cal H}$.
In this case, the DDMMC algorithm randomly picks a charge, and then randomly picks a direction
in which to move it.  The move will be accepted only if it lowers the energy,
i.e. if the charge advances in the direction of ${\bf F}$.  This will happen only for a
certain fraction $p$ of the possible directions.  For a triangular grid, with ${\bf F}$ 
aligned with one of the grid basis vectors, $3$ of the $6$ possible nearest neighbor 
directions will have a component parallel to ${\bf F}$ and so $p=1/2$; for a square
grid, $p=1/4$.  Thus, after one pass through the system, a randomly selected fraction $p$
of the charges have advanced forward, while the rest remain in place.  After a second
pass through the system, a different randomly selected fraction $p$ move forward.
After many such passes one expects the system to be disordered.  In fact, we find
from simulations that, at $T=0$, DDMMC disorders the ground state charge lattice 
for {\it all} $F>F_c$.  The randomness of choosing moves, inherent in the
DDMMC algorithm, is sufficient to disorder the moving system even as $T\to 0$.

\begin{figure}
\epsfxsize=8.truecm
\epsfbox{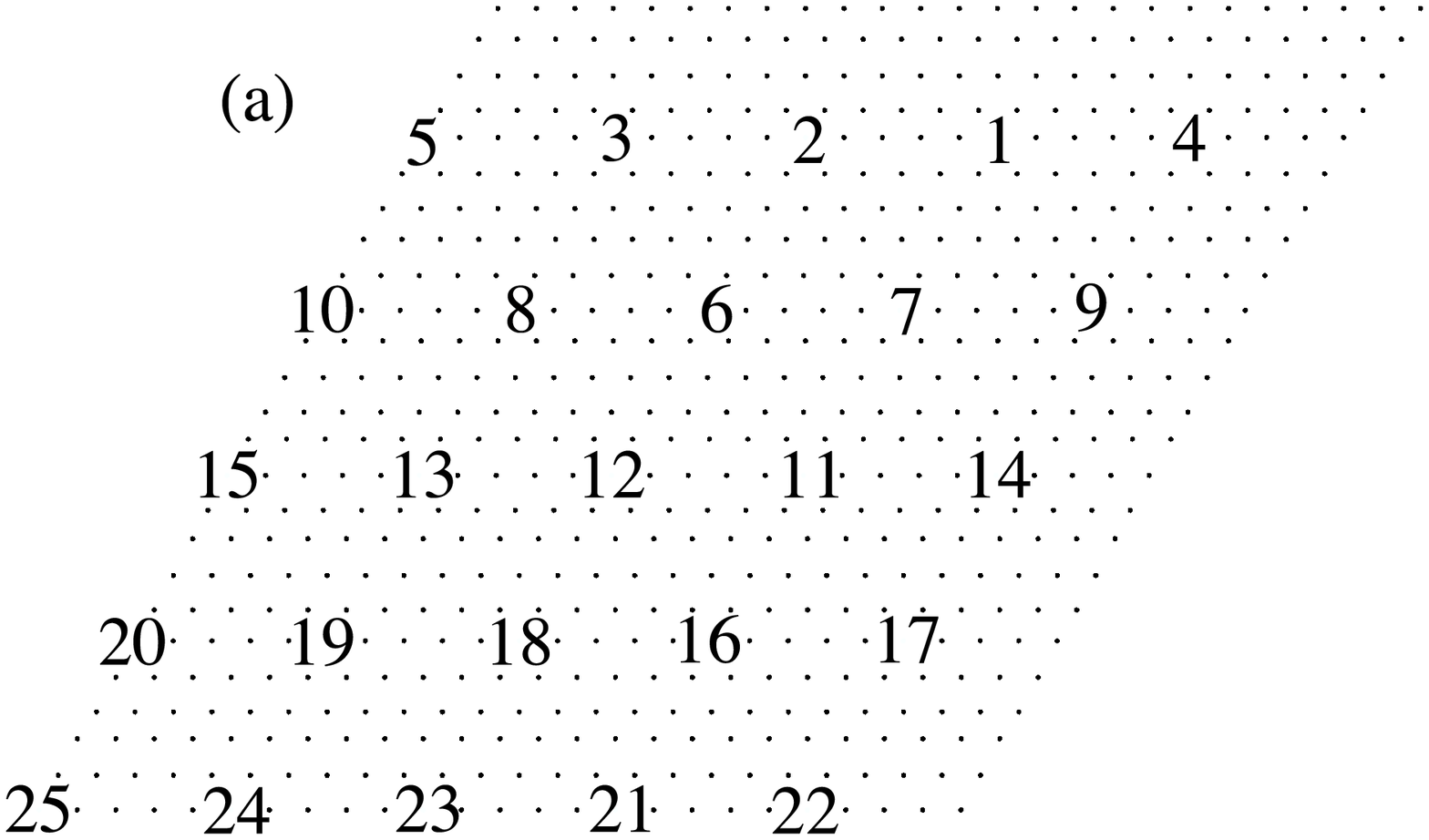}

\epsfxsize=8.6truecm
\epsfbox{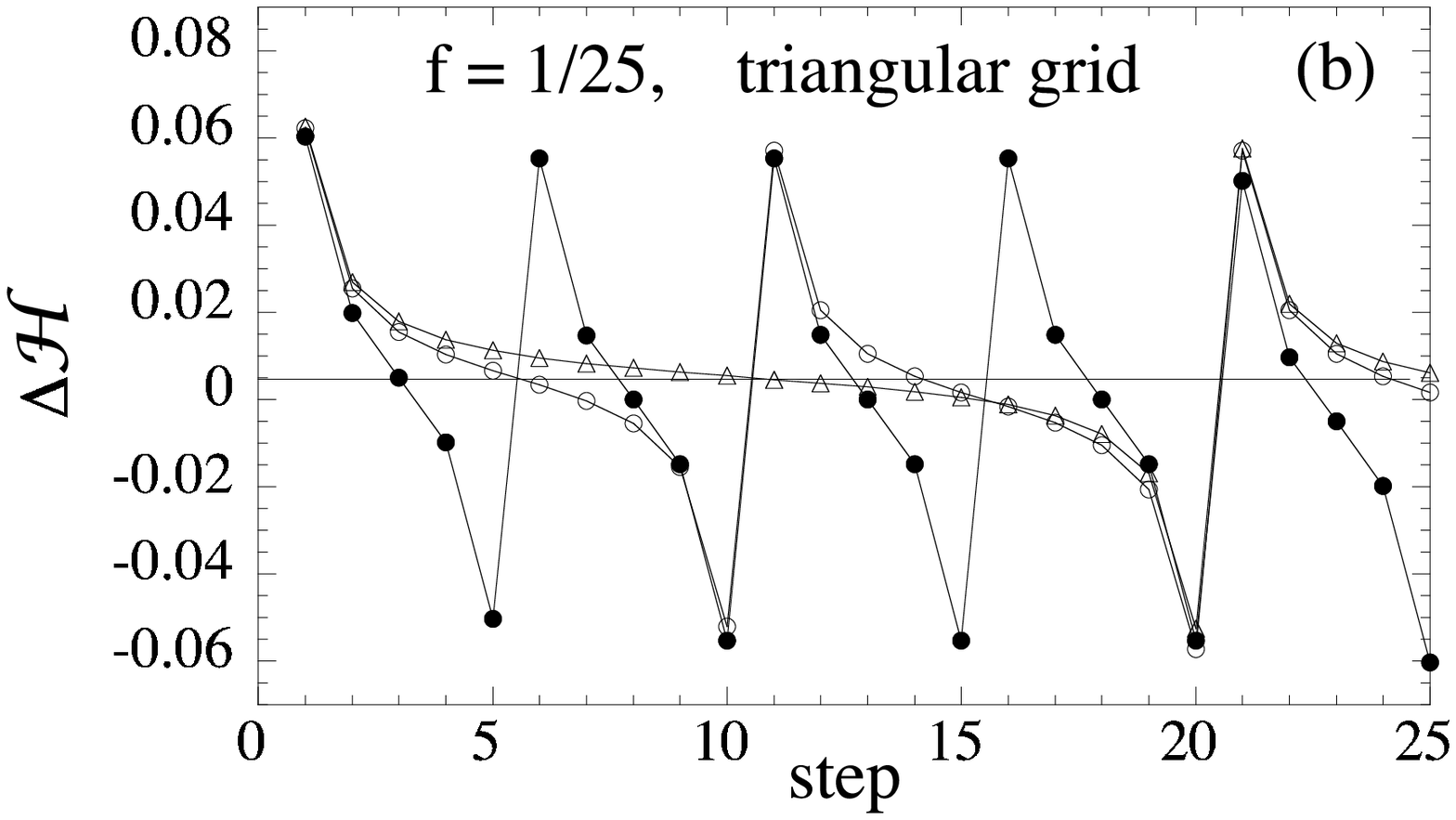}
\caption{CTMC on a triangular grid with charge density $f=1/25$
at $T\to 0$ and $F>F_c$, with ${\bf F}$ parallel to the $\hat a_1$ axis.
(a) Ground state charge lattice for a $25\times 25$ triangular grid. 
Numbers denote the locations of the charges in the
ground state.  The value of each number indicates the step on which that charge
moves forward.   (b) The change in interaction
energy $\Delta{\cal H}$ at each step as charges move forward. ($\bullet$) are for
a $25\times 25$ grid and correspond to the moves in (a); ($\circ$) and ({\scriptsize $\triangle$})
are the beginnings of similar sequences for $50\times 50$ and $100\times 100$ grids.
}
\label{f2}
\end{figure}

We now consider the low $T$ behavior of CTMC.  For specificity we will
consider the case of a triangular grid with a charge density of $f=1/25$,
and ${\bf F}=F\hat a_1$ aligned along one of the grid axes.
We will study this particular case in great detail in section \ref{sResultsTri}.
Consider the limit $T\to 0$, starting in the $F=0$ ground state and then increasing $F$,
but with $F<F_c$.  
The configuration of charges in the $F=0$ ground state 
is shown in Fig.\,\ref{f2}a for a $25\times 25$ grid.
Since CTMC is a rejectionless algorithm, even when $F<F_c$ CTMC will
make an excitation out of the ground state.  However since $\Delta U>0$
for this excitation, the time $\Delta t$ for this excitation to occur diverges
exponentially as $T\to 0$.  Conversely, once an excitation has been made,
the very next move will be to relax the excitation back to the ground state,
since this is the only move for which $\Delta U<0$; moreover, since $\Delta U<0$,
this move takes place in an exponentially vanishing time.  Thus alternating steps
of CTMC will consist of displacing a randomly selected single charge and 
then moving it back.  As $T\to 0$,
the simulation clock time to be in the ground state grows exponentially large, while
the clock time to be in the excited state gets exponentially small.
The system therefore remains pinned in the ground state, with the time in the
excited states contributing negligibly to any measured averages.

Next, consider what happens when $F>F_c$.  Because of the
rates of Eq.\,(\ref{eWialpha}), as $T\to 0$ only moves with
the smallest value of $\Delta U_{i\alpha}\equiv \Delta U_{\rm min}$ can be selected; all
other possible moves are exponentially suppressed.  This results
in the main difference between CTMC and DDMMC.  In CTMC,
as $T\to 0$, motion is {\it deterministic} except for choosing randomly
among moves with degenerate values of $\Delta U_{\rm min}$.
Now consider starting in the $F=0$ ground state
shown in Fig.\,\ref{f2}a.  All moves that advance a charge forward one
grid spacing along $\hat a_1$ are equally likely, with $\Delta U_{{\rm min}, 1} = F_c-F<0$, 
while moves in all other directions are exponentially suppressed.  
Thus the first step will be to select any one of the $N_c$ charges at random and move it 
forward.  On the second step, however, there are only two moves that have
the new lowest $\Delta U_{{\rm min}, 2}$; these are to advance 
either the charge immediately in front of, or the charge immediately behind, the charge that
moved in the first step.  On the third step there are similarly only two moves
with $\Delta U_{{\rm min}, 3}$; advancing the charge immediately in front of,
or immediately behind, the first two moved charges.  The system proceeds
in a similar manner until all charges in the same row parallel to ${\bf F}$
have moved forward.  The next move will be to pick a charge at
random in one of the two adjacent charge filled rows and move it forward, and then
subsequently all charges in that row move forward one by one.  In this manner,
the rows of charges move one after another forward until the system has returned to the
starting ground state, but with the entire charge lattice advanced by one grid
spacing.  In Fig.\,\ref{f2}a we label each of the ground state charges by the step number
on which that charge moved forward in one particular pass of CTMC on a $25\times 25$
size grid.  The pattern described above is clearly evident.  In Fig.\,\ref{f2}b we
plot the change in interaction energy, $\Delta{\cal H}=\Delta U_{{\rm min},n}+F$, for 
each step $n$ of this pass; note that $\Delta{\cal H}$ is independent of the
applied force $F$.  The rough oscillation of $\Delta{\cal H}$ with a period of $n=L_1/a_0$,
with $a_0=1/\sqrt{f}$ the spacing between the charges, reflects the row by row
motion of the charges.

Next we consider the timing of the above sequence of moves.  From Fig.\,\ref{f2}b
we conclude that for each step $n>1$ of the above pass, $U_{{\rm min},n}<U_{{\rm min},1}$.
Hence  the rate, Eq.\,(\ref{eWialpha}), to make any step $n>1$ is
exponentially larger than the rate to make the first step, $n=1$.
As $T\to 0$ we conclude that the relative time spent in the 
intermediate states (i.e. the states after steps $n=1\dots N_c-1$) as compared to
the time spent in the ordered ground state (prior to the first step and after step $n=N_c$) 
vanishes exponentially.  According to the simulation clock time, all charges in the ground 
state charge lattice have advanced forward one grid spacing essentially {\it simultaneously}.
This is the deterministic motion of the charge lattice that one would physically 
expect to find for $F>F_c$ as $T\to 0$.

There is, however, one peculiar aspect to the above $T\to 0$ dynamics.  By the above
arguments, the velocity of the moving charge lattice will be proportional to the
rate to make the initial first step.  From Eqs.\,(\ref{eWialpha}) and (\ref{eDt}) this rate will be
$W_{\rm tot}=N_cW_0{\rm e}^{-\Delta U_{{\rm min},1}/2T}$.  Thus the $T\to 0$
velocity grows proportional to the number of charges $N_c$ in the system.
However this can be understood physically if one views motion on
the discrete grid as being a representation for continuum motion in a periodic potential.
In this case, one should take $W_0\sim {\rm e}^{-2U_0/T}$, as in Eq.\,(\ref{eWAHs}),
where $2U_0$ is the maximum to minimum energy difference of the potential.
In the term $W_{\rm tot}$ above, the factor $W_0$ represents the rate for
a particular charge to be excited out of the ground state, over the energy barrier $2U_0$
into the neighboring down stream potential minimum, thus lowering the energy of the system by
$\Delta U_{{\rm min},1}$.  This rate becomes exponentially slow as $T\to 0$.
Once this initial excitation has taken place, all the other charges follow, advancing
forward in what may be regarded as an ``avalanche".  We call this an avalanche
because all the other charges move forward in a time that becomes vanishingly small 
compared to the time to make the initial excitation.  The factor $N_c$ in $W_{\rm tot}$
just reflects the $N_c$ possible sites at which the initial
excitation that leads to the avalanche can occur.  Thus, while $W_{\rm tot}$ grows
as $N_c$, it also vanishes exponentially as $T\to 0$, and so at $T=0$ the charges are always
pinned, as is physically appropriate for $F_c<F<U_0$.

Several features of the expected behavior at {\it finite} $T$ can also be inferred
from Fig.\,\ref{f2}b.  Once a first charge in a given row has moved forward,
the energy change for the other charges in the same row to move forward rapidly
decreases.  Consequently, once the first charge has moved forward, the remaining
charges in that row rapidly follow forward.  However, once a row has completely
moved forward, the energy for the first charge in an adjacent row to move forward
is not much lower than the energy for a first charge in any other row to move forward.
Comparing the values of $\Delta{\cal H}$ in Fig.\,\ref{f2}b for steps $n=1$ and
$n=(L_1/a_0)+1$, we estimate this energy difference as $\Delta E\simeq 0.008$.
We therefore expect that once the temperature $T$ become of the same order as
this $\Delta E$, coherence between moving rows will be lost.  Avalanches will
now consist of individual rows moving forward, but different rows will be 
uncorrelated.  Consequently,
the average velocity in this regime will scale proportionally to $L_1/a_0$ (the number
of sites in a given row for an initial excitation to occur) rather than $N_c$.
The details of this picture will depend on the specific correlations
between charges within a given row, versus between rows, and this will
be a subject of investigation in section \ref{sResultsTri}.

\subsection{Observables}

To determine the properties of our non-equilibrium steady states, we 
measure several static (time independent) and dynamic (time dependent)
quantities.  To determine structural properties, the main quantity of interest
is the structure function,
\begin{equation}
S({\bf k})\equiv{1\over N_c}\langle n_{\bf k}n_{-{\bf k}}\rangle\enspace,
\label{eSk}
\end{equation}
where,
\begin{equation}
n_{\bf k}=\sum_i{\rm e}^{i{\bf k}\cdot{\bf r}_i}n_i\enspace,
\label{enk}
\end{equation}
is the Fourier transform of the charge distribution $n_i$, and ${\bf k}$ is
one of the allowed wavevectors in the first Brillouin zone (shown in 
Fig.\,\ref{f1}b for the triangular grid).  
The corresponding real space correlation function is given by,
\begin{equation}
C(m_1,m_2)\equiv{1\over N}\sum_{k_1,k_2}{\rm e}^{-2\pi i(m_1k_1+m_2k_2)}S(k_1,k_2)
\enspace,
\label{eC}
\end{equation}
where we have expressed the positions ${\bf r}_i$ and wavevectors ${\bf k}$
in terms of their coordinates $m_\mu$ and $k_\mu$, as in Eqs.\,(\ref{eri}) and
(\ref{ek}), in constructing the above Fourier transform.  We will also consider
the mixed correlation,
\begin{equation}C(k_1,m_2)\equiv{1\over N}\sum_{k_2}{\rm e}^{-2\pi i m_2k_2}S(k_1,k_2)
\enspace.
\label{eCmixed}
\end{equation}

The above quantities give information about the {\it translational} order of the system.
To investigate the {\it orientational} order, we define the $6$-fold orientational (hexatic)
order parameter,
\begin{equation}
\Phi_6\equiv {1\over N_c}\sum_i{1\over z_i}\sum_j {\rm e}^{6i\theta_{ij}}\enspace.
\label{ePhi6}
\end{equation}
In the above, the first sum is over all charges $n_i=1$, the second sum is over all
charges $n_j=1$ that are nearest neighbors of $n_i$, $z_i$ is the number of
such nearest neighbors, and $\theta_{ij}$ is the angle of the bond from
$n_i$ to $n_j$ with respect to the $\hat a_1$ axis.  Nearest neighbors are
determined by Delaunay triangulation \cite{Delaunay}.

We also measure several dynamical quantities.  Let
\begin{equation}
{\bf R}_{\rm cm}(t)\equiv {1\over N_c}\sum_{s<t}\Delta {\bf r}_s
\label{ecm}
\end{equation}
be the net displacement of the center of mass of the charges
at time $t$ of the simulation clock.  The right hand side of the
above is just the sum of charge displacements at each step $s$
of the simulation that occurs before the simulation clock has
reached time $t$, normalized by the total number of charges.
The average velocity of the system is then just,
\begin{equation}
{\bf v}_{\rm ave} = {{\bf R}_{\rm cm}(\tau)-{\bf R}_{\rm cm}(t_{\rm eq})\over \tau - t_{\rm eq}}\enspace,
\label{evave}
\end{equation}
where $\tau$ is the total simulation clock time, and $t_{\rm eq}$ is some
initial time to allow for equilibration.  In the analogy between 2D charges
and vortices in a superconducting film, the average charge velocity becomes
the average voltage drop transverse to the direction of motion of the vortices.

We also look at the fluctuations about the average center of mass position.
If we define the fluctuation, after a time $t$, about the average center of mass position,
\begin{equation}
\delta {\bf R}(t;t_0)\equiv {\bf R}_{\rm cm}(t+t_0)-{\bf R}_{\rm cm}(t_0)-
{\bf v}_{\rm ave}t\enspace,
\label{edR}
\end{equation}
then we can define the diffusion tensor ${\bf D}(t)$ by,
\begin{equation}
2{\bf D}(t)t\equiv N_c \langle \delta{\bf R}(t;t_0)\delta{\bf R}(t;t_0)\rangle_{t_0}\enspace,
\label{eDtensor}
\end{equation}
where the angular brackets in the above denote an average over the parameter $t_0$
during the course of a single simulation.  In averaging over $t_0$, we restrict ourselves
to non-overlapping intervals, i.e. to the values $t_0=nt$, for integer $n$, so as to reduce 
correlations among the different terms being averaged.  If fluctuations about the
center of mass are diffusive, then we expect ${\bf D}(t)$ to saturate to a constant
as $t$ increases.  The factor $N_c$ on the right hand side of Eq.\,(\ref{eDtensor})
ensures that ${\bf D}(t)$ approaches a size independent value in the liquid state,
where the charges have only short ranged correlations.

Although in our simulation we will use Eq.\,(\ref{eDtensor}) to compute ${\bf D}(t)$,
the diffusion tensor can also be expressed in its more familiar form, in terms of velocity correlations \cite{Minnhagen}.  If we define the instantaneous fluctuation in velocity by,
\begin{equation}
\delta {\bf v}(t)\equiv {\bf v}(t)-{\bf v}_{\rm ave}=\delta {\bf R}(\Delta t;t)/\Delta t
\enspace,
\label{edvel}
\end{equation}
then
\begin{equation}
\lim_{t\to\infty} {\bf D}(t)={N_c\over 2}\int_{-\infty}^{\infty}dt\langle\delta {\bf v}(t)\delta {\bf v}(0)\rangle\enspace.
\label{eDtensor2}
\end{equation}
For a superconducting network, where vortex velocity is proportional to the
voltage drop in the direction transverse to the vortex motion, ${\bf D}$ is
a measure of the voltage fluctuations.

In equilibrium, when ${\bf F}={\bf v}_{\rm ave}=0$, ${\bf D}/T$ is proportional to 
the charge mobility tensor by the fluctuation-dissipation theorem \cite{flucDis}.  In the analogy
to vortices in superconducting films, this is the linear resistivity of the film.  In the driven
state, with ${\bf F}=F\hat x$, we will use the transverse component of the 
diffusion tensor, $D_{yy}$, to test for the presence of {\it transverse} pinning.
If $D_{yy}>0$, the center of mass is diffusing transversely to the direction of
the average motion; application of a small transverse force $\delta F\hat y$ will
cause the system to acquire a transverse component of the velocity, $v_y\propto
\delta F$.  In analogy with equilibrium, we will assume that if $D_{yy}=0$,
there will be no linear transverse response, i.e. $v_y/\delta F\to 0$ as $\delta F\to 0$.
This characterizes a {\it transversely pinned} state \cite{Giamarchi}.



In CTMC, averages in the steady state are computed by the time integral in
Eq.\,(\ref{eO}).  However, the direct application of Eq.\,(\ref{eO}) would require
the evaluation of the measured quantity after {\it every} single step of the simulation.
For quantities involving lengthy calculation, such as $S({\bf k})$ and $\Phi_6$, 
this is not practical except for fairly
short runs.  Instead, we compute the time integrals for these quantities by a 
Monte Carlo integration \cite{NumRec}, averaging them over $N_{\rm config}$
configurations sampled randomly, with a uniform distribution, over the simulation clock time
interval $(t_{\rm eq},\tau)$, with $t_{\rm eq}$ an initial equilibration time and
$\tau$ the total simulation time.  In practice, we implement this scheme as follows.
We compute the average time interval between samplings, $\tau^\prime=(\tau-t_{\rm eq})
/N_{\rm config}$.  Then, after a first sampling, we determine the time until the
next sampling by drawing from an exponential distribution with average time constant
$\tau^\prime$.  This gives the correct sampling since, if $t_1<t_2\dots <t_n$ are
the ordered values of $n$ independent and uniformly distributed random variables
on a given interval, the probability distribution for the distance $t_{i+1}-t_i$
is exponential.  We use typically $N_{\rm config}\simeq 10^3$ to $10^4$ in our simulations.

\section{Results on a triangular grid}
\label{sResultsTri}

We now report our results for the case of charges on a triangular grid. 
The {\it equilibrium}, i.e. ${\bf F}=0$, behavior  \cite{Franz} of this system depends
on the charge density $f$.  For sufficiently dense $f$ (but not too dense), 
there is only a
single first order melting transition at $T_{\rm m}$, from a pinned charge
solid with long range translational order at low $T$, to an ordinary liquid
at high $T$.  For more dilute $f$, there are three phases: a low $T$ pinned
solid with long range translational order, an intermediate $T$ floating solid
with algebraic translational order, and a high $T$ liquid.  In this work we
will consider the charge density $f=1/25$, which falls in the dense limit
with a single first order equilibrium melting temperature of $T_{\rm m}\simeq 0.0085$.
The dilute limit will be considered elsewhere.

\subsection{Phase Diagram}

\begin{figure}
\epsfxsize=8.6truecm
\epsfbox{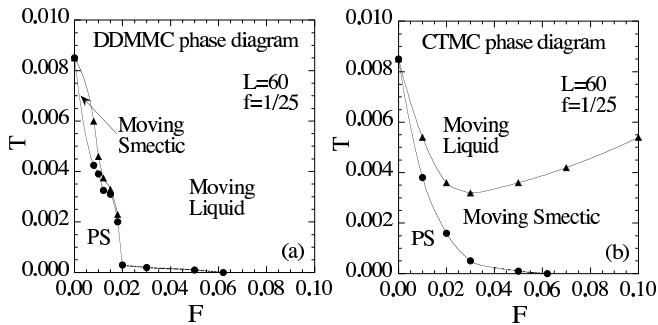}
\caption{Phase diagram for a $60\times 60$ triangular grid with charge density
$f=1/25$ as a function of temperature $T$ and driving force ${\bf F}=F\hat x$,
(a) for DDMMC dynamics, and (b) for CTMC dynamics.  ``PS" stands for pinned solid.
}
\label{f3}
\end{figure}

We start by mapping out the $T-F$ phase diagram for a $60\times 60$
grid, with the applied force along the $\hat a_1$ grid axis, ${\bf F}=F\hat x$.
We initialize the system in the ordered $F=0$ ground state, set $F$ to
the desired driving force, and then simulate the system for increasing values of $T$.
By measuring the average interaction energy $\langle{\cal H}\rangle$, 
structural properties such as $S({\bf k})$ and $\langle\Phi_6\rangle$, and
the average velocity ${\bf v}_{\rm ave}$, we determine the phase diagrams
shown in Fig.\,\ref{f3}.  Our simulations consist typically of $\sim 10^5$ to $10^6$ passes
through the system, depending on system size and parameters.  

Our results for the DDMMC dynamics are shown in Fig.\,\ref{f3}a.
As discussed earlier in section \ref{sLowT}, at $T=0$ the system remains
pinned (${\bf v}_{\rm ave}=0$) in its equilibrium ground state for all $F$ 
below a critical force $F_c=0.0603$; for all $F>F_c$ the moving state is
a liquid \cite{Gotcheva}.  For fixed $F<F_c$, upon increasing $T$, the solid remains
pinned with long range translational order, up until a value $T_{\rm p}(F)$.
It then enters a moving state with 
finite ${\bf v}_{\rm ave}$.  Over most of the phase
diagram we find \cite{Gotcheva} that this moving state is a liquid with 
a correlation length of order the spacing between charges $a_0$.
Only in a very narrow region at low $F$ and
higher $T$ do we find a structure that appears to be a moving smectic
phase.  We will discuss what we mean by a ``smectic" phase in greater
detail when we describe our results from CTMC.  For DDMMC we have
not investigated in any detail the stability of this small region of smectic
phase with respect to increasing the system size, or with respect to
cooling from the liquid.
The lack of structure for almost all of the moving state, particularly at large
$F$ and small $T$, suggests that the DDMMC algorithm is indeed
unphysical and unlikely to be a good model for continuum dynamics.
We therefore focus on the CTMC algorithm for the remainder of this paper.

\begin{figure*}
\epsfxsize=17.2truecm
\epsfbox{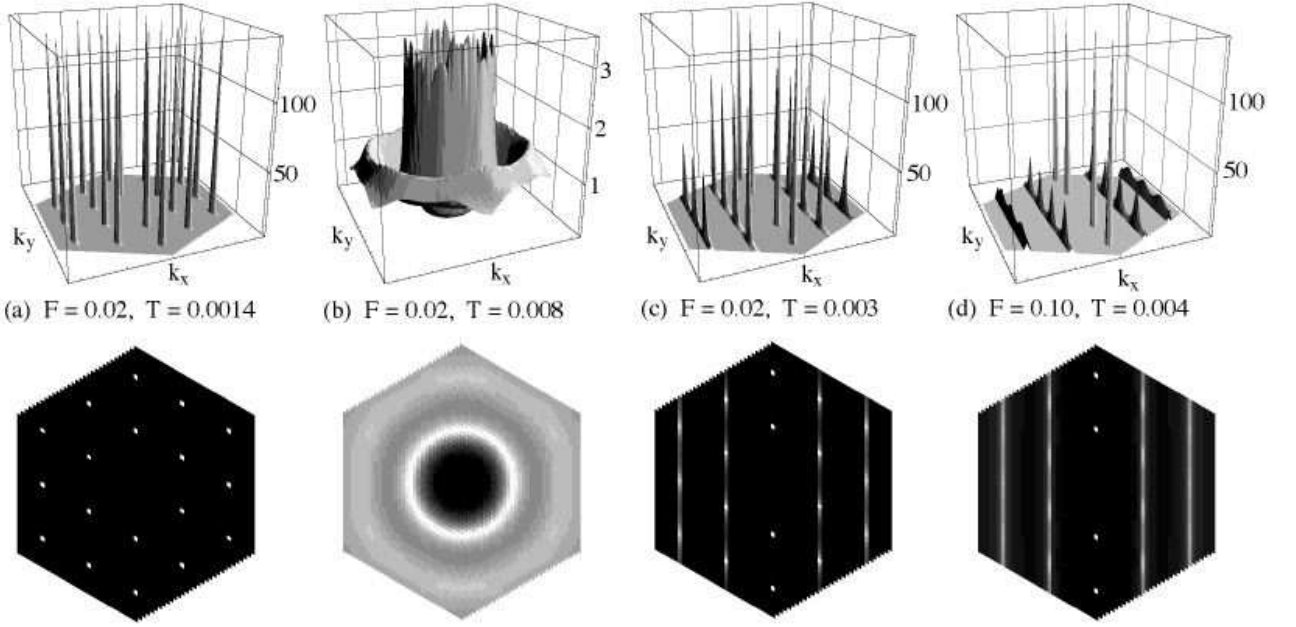}
\caption{S({\bf k}) for several points in the CTMC phase diagram of
Fig.\,\protect\ref{f3}b. (a) $F=0.02$, $T=0.0014$ in the pinned solid;
(b) $F0.02$, $T=0.008$ in the moving liquid; (c) $F=0.02$, $T=0.003$
in the moving smectic; (d) $F=0.10$, $T=0.004$ in the moving smectic at
higher drive.   The bottom row shows intensity plots of the corresponding
graphs in the top row.  The peak $S({\bf k}=0)=N_c$ is removed to give
greater contrast to the other peaks.
}
\label{f4}
\end{figure*}

In Fig.\,\ref{f3}b we give the phase diagram for CTMC dynamics.  
Again we find pinned, liquid, and smectic phases, but now the smectic
persists over a wide region of the $T-F$ plane, particularly at low $T$
and large $F$.  To illustrate the nature of order in each of these phases, we
plot in Fig.\,\ref{f4} the structure function $S({\bf k})$ for several
representative points in the phase diagram.  In Fig.\,\ref{f4}a we see
the sharp Bragg peaks with $S({\bf K})\simeq N_c$, 
characteristic of the long range translational order
in the pinned solid phase.  The peaks are at the reciprocal lattice vectors
of the ordered charge solid, and given by,
\begin{eqnarray}
&&{\bf K}_{p_1,p_2}=k_1{\bf b}_1+k_2{\bf b}_2\nonumber\\
&&{\rm with}\quad k_\mu = {p_\mu\over 5},\quad p_\mu=0, \pm 1,\pm 2\enspace.
\label{eKK}
\end{eqnarray}
Fig.\,\ref{f4}b shows a roughly circular and
finite peak, characteristic of short range translational order in the
moving liquid phase.  

Figs.\,\ref{f4}c,d show the moving smectic at
small and large driving forces, respectively.  Consider first the large drive
case in  Fig.\,\ref{f4}d.  The peaks along the $k_2$ axis ($k_1=0$) at $(k_1,k_2)=(0,\pm 1/5)$
and $(0,\pm 2/5)$ (see Fig.\,\ref{f1}b for the $k$-space geometry) 
are sharp Bragg peaks with $S({\bf K})\simeq N_c$.
This indicates that if one averages the particle density in the $\hat a_1$ direction
($k_1=0$), the resulting density is periodic in the $\hat a_2$ direction with a
period of $5$ grid spacings;
the particles are therefore moving in periodically spaced channels 
oriented parallel to the driving force ${\bf F}=F\hat a_1$.
Next, note that the peaks at finite $k_1=\pm 1/5,\pm 2/5$  appear to be
sharp, i.e. only one grid spacing wide, along the $k_1$ direction.
Such $\delta$-function like peaks in the $k_1$ direction
indicate that the particles are periodically ordered within each smectic channel
with a period of $5$ grid spacings.
The finite width of these peaks in the $k_2$ direction indicates that the
ordered smectic channels are randomly displaced with respect to each other, with
a finite correlation length $\xi_\perp$ proportional to the inverse width of the peak.
Comparing Fig.\,\ref{f4}c with Fig.\,\ref{f4}d, we see similar
features at the smaller drive $F$, 
only the peaks at finite $k_1$ are now sharper, with a narrower width 
in the $k_2$ direction.  In the next two sections we will consider these features
of the smectic phase in much greater detail, studying the scaling behavior and
stability of the smectic as the system size increases.

\begin{figure}
\epsfxsize=8.6truecm
\epsfbox{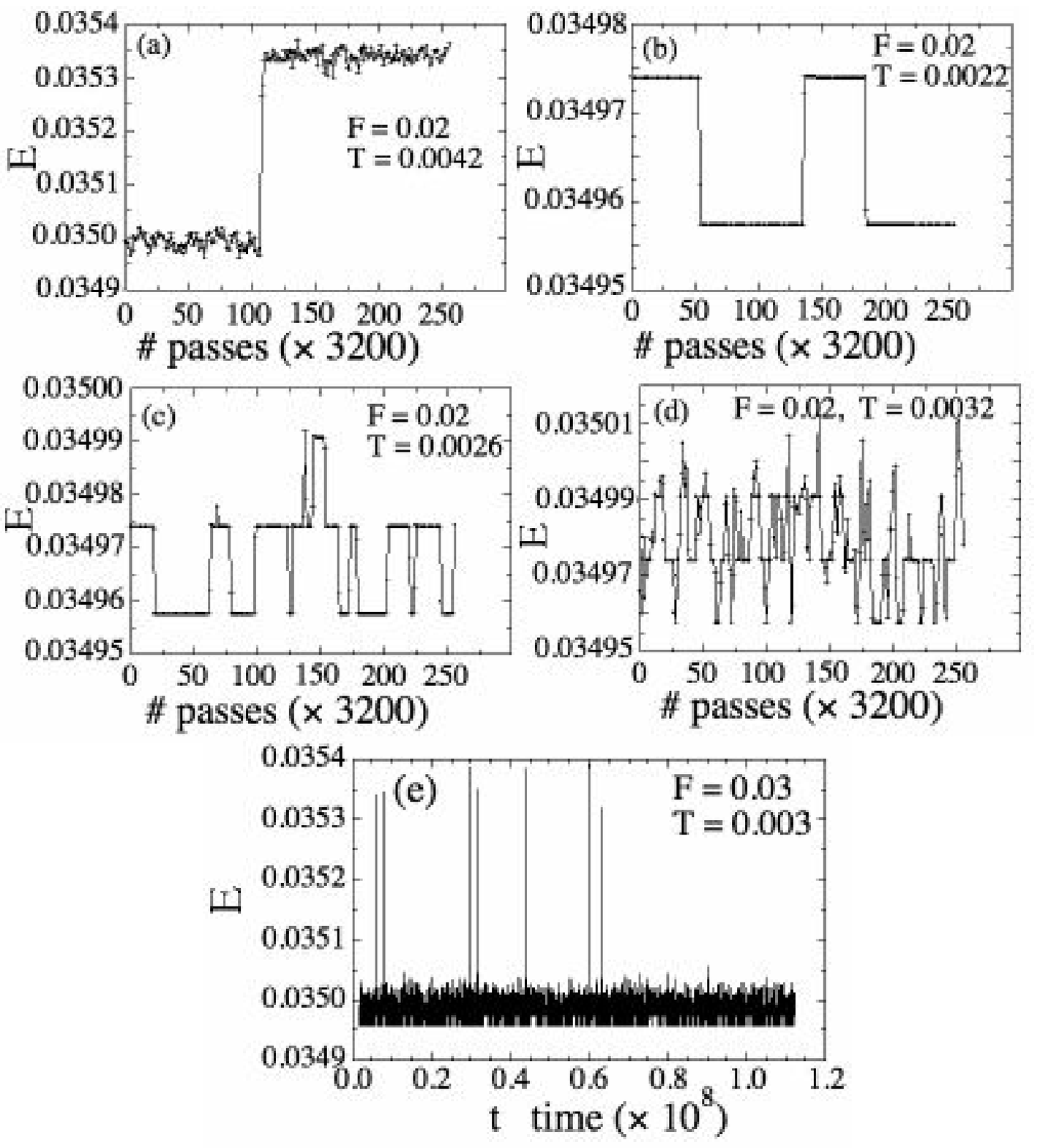}
\caption{(a)-(d): Average interaction energy per site $E=\langle{\cal H}/N\rangle$
vs. the number of simulation passes through the grid.  Each point represents
an average over $3200$ successive passes; (a) $F=0.02$, $T=0.0042$, just
at the melting transition $T_{\rm m}(F)$; (b) $F=0.02$, $T=0.0022$, just 
above the unpinning transition $T_{\rm p}(F)$; (c) $F=0.02$, $T=0.0026$ and
(d) $F=0.02$, $T=0.0032$, moving away from the unpinning transition.
(e) E vs. simulation clock time $t$, for $F=0.03$, $T=0.003$, slightly below
the melting transition.
}
\label{f5}
\end{figure}

Finally we consider the nature of the melting transition $T_{\rm m}(F)$ from the
smectic to the  liquid, and the unpinning transition $T_{\rm p}(F)$ from the pinned
solid to the smectic.  In Fig.\,\ref{f5}a$-$d we plot the average interaction energy
per site, $E=\langle{\cal H}/N\rangle$ vs. the number of simulation passes through
the grid.  Each point represents an average over $3200$ successive passes.  
Fig.\,\ref{f5}a shows results at $F=0.02$, $T=0.0042$, just at the melting
transition $T_{\rm m}(F)$.  We see that the energy takes
a discontinuous jump as the system makes the transition from smectic to liquid.
Melting of the smectic is therefore like a first order phase transition.
Fig.\,\ref{f5}b, shows results at $F=0.02$, $T=0.0022$, just above the unpinning
transition $T_{\rm p}(F)$.  We see that the energy fluctuations form a set of plateaus, 
with a very long period of fluctuation.  The lowest plateau corresponds to the 
ordered $F=0$ ground state with a value $E_0=0.03495736$.  The higher energy
plateau corresponds to having some fraction of adjacent smectic channels (i.e.
charge filled rows) advanced one
grid spacing parallel to ${\bf F}$, so that the system looks locally like the ground state, 
but with one pair of domain walls parallel to ${\bf F}$.  

As $T$ increases above $T_{\rm p}$, Figs.\,\ref{f5}c,d
show that the rate of fluctuations increase, and plateaus of additional energy values appear.
The higher energy plateaus correspond to having more than one pair of domain walls
in the otherwise ordered system. This behavior may be understood by considering the results
shown in Fig.\,\ref{f2}b.  For a driving force of $F=0.02$, the thermal
energy needed to excite a pair of adjacent particles in a given row to move forward is 
$\Delta U=\Delta U_{{\rm min},1}+\Delta U_{{\rm min},2}
\approx 0.048$; however the energy to move each remaining particle is 
$\Delta U_{{\rm min},n}<0$.
Thus, at low $T$, the excitation of an initial pair forward will trigger the remaining particles
in that row to move forward almost instantaneously.  
The rate for the initial pair excitation goes as, $W\sim {\rm e}^{-\Delta U/2T}$,
vanishing exponentially as $T$ decreases.  The rate of energy fluctuations 
decreases accordingly.  At low $T$, once a given row has moved
forward, the next most favorable excitation is to move an adjacent row of charges forward
(see discussion at the end of section \ref{sLowT}).  The system thus consists of 
a single pair of domain walls in the otherwise ordered ground state; the distance
between the domain walls increases as more adjacent rows move forward.  Such states give
the higher of the two energy plateaus in Fig.\,\ref{f5}b.  As $T$ increases, there becomes
a non-negligible probability to have a pair excitation in a non-adjacent row of charges.  Now
the system can develop more than a single pair of domain walls, leading to the
additional high energy plateaus of Figs.\,\ref{f5}c,d.  

The above arguments 
suggest that $T_{\rm p}(F)$ may not be a true phase transition.  Since the above rate 
$W$ is finite at any $T$, but grows vanishingly small as $T\to 0$, the observed
$T_{\rm p}(F)$ may just result from $1/W$ growing larger than the longest
simulation time we can carry out.  As $F$ decreases, it will becomes necessary to 
excite three, then four, then more, particles forward in a given row, before one
reaches the condition that $\Delta U_{{\rm min},n}<0$ triggering the remaining
particles in the row to move immediately forward (see Fig.\,\ref{f2}b).  Thus we expect that 
$W$ will decrease, and the observed $T_{\rm p}(F)$ will increase, as $F\to 0$.
Note that the graphs in Fig.\,\ref{f5}a$-$d are plotted versus the number of 
simulation passes rather than the simulation clock time.  They therefore reflect the
amount of actual computation needed to observe fluctuations of the system.
The decrease in fluctuation rate observed as $T$ decreases for fixed $F=0.02$
(compare Fig.\,\ref{f5}d to Fig.\,\ref{f5}b)
results from the decrease in probability to move the second particle of an excitation
pair forward, {\it before} the first particle has had a chance to fall back into place,
rather than being directly due to the overall exponential decrease with $T$ of all single
particle rates as in Eq.\,(\ref{eWialpha}).

Finally we note that, for finite system size, the moving smectic state is the
true stable steady state of the system.  In Fig.\,\ref{f5}e we plot the average
interaction energy per site $E$ versus simulation clock time $t$, for the parameters
$F=0.03$, $T=0.003$, which lies just immediately below the melting transition
$T_{\rm m}(F)$ (see Fig.\,\ref{f3}b).  Comparing with Fig.\,\ref{f5}a, we see clearly
that the system has occasional fluctuations into the liquid phase, as indicated by the
large brief spikes in energy. Such fluctuations are expected for a finite size system
near a first order phase transition.  The fact that the system returns to the smectic
state, after such a liquid fluctuation, indicates that the smectic is indeed the stable
steady state.  We have also succeeded in cooling into the smectic state from the
disordered liquid, and in entering the smectic from the liquid by increasing $F$
at temperatures above the minimum of the $T_{\rm m}(F)$ transition boundary.

\subsection{Smectic Phase - High Drive}
\label{sHigh}

\begin{figure}
\epsfxsize=7.5truecm
\epsfbox{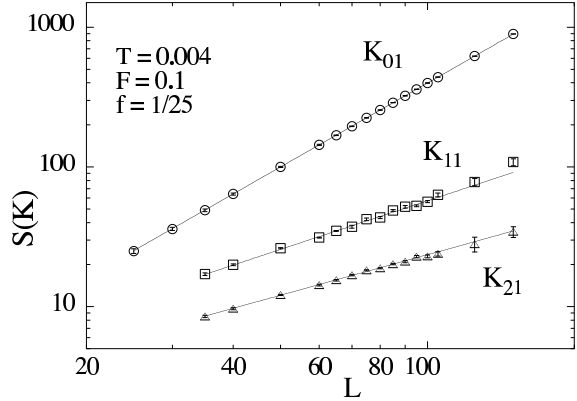}
\caption{Scaling of peak heights $S({\bf K})$ vs. system size $L$
for the smectic phase at $F=0.10$, $T=0.004$.  Straight lines indicate
good power law fits, $S({\bf K})\sim L^s$, with $s\simeq 1.99$ for ${\bf K}_{01}$,
$s\simeq 1.15$ for ${\bf K}_{11}$, and $s\simeq 0.96$ for ${\bf K}_{21}$.
}
\label{f6}
\end{figure}

In the next two sections we explore in detail the nature of ordering in the moving smectic
state, and its stability as the system size increases.   We start here by considering the
smectic in the high drive case at $F=0.1$, $T=0.004$, corresponding to Fig.\,\ref{f4}d.
If the system has true long range smectic order, we expect the peaks in $S({\bf k})$ along
the $k_2$ axis ($k_1=0$) to be true Bragg peaks, with a height that scales as
the system area.  In Fig.\,\ref{f6} we plot the height of the peak $S({\bf K}_{01})$,
versus system length $L$, for systems of size $L\times L$.  The straight line
on the log-log plot has a slope $s\simeq 1.99$ giving good agreement with the
$\sim L^2$ behavior expected for long range smectic order.

Next we consider the ordering within the smectic channels.  If charges have long
range order within each individual channel, we expect the peaks in $S({\bf k})$ at 
$k_1=\pm 1/5,\pm 2/5$ to be $\delta$-function like in the $k_1$
direction.  If the channels have only short range correlations between them, the 
width of these peaks will remain finite in the $k_2$ direction.  We therefore
expect that the heights of these peaks at finite $k_1$ should scale as the
system length $L_1$ in the $\hat a_1$
direction.  In Fig.\,\ref{f6} we plot the height of two of these peaks, 
$S({\bf K}_{11})$ and $S({\bf K}_{21})$
(see Eq.\,(\ref{eKK}) for our notation labeling the reciprocal lattice vectors ${\bf K}$)
versus system length $L$, for systems of size $L\times L$.  The straight lines
have slopes $s=1.15$ and $0.96$ respectively, in reasonable agreement with the
$\sim L$ behavior described above.

\begin{figure}
\epsfxsize=7.5truecm
\epsfbox{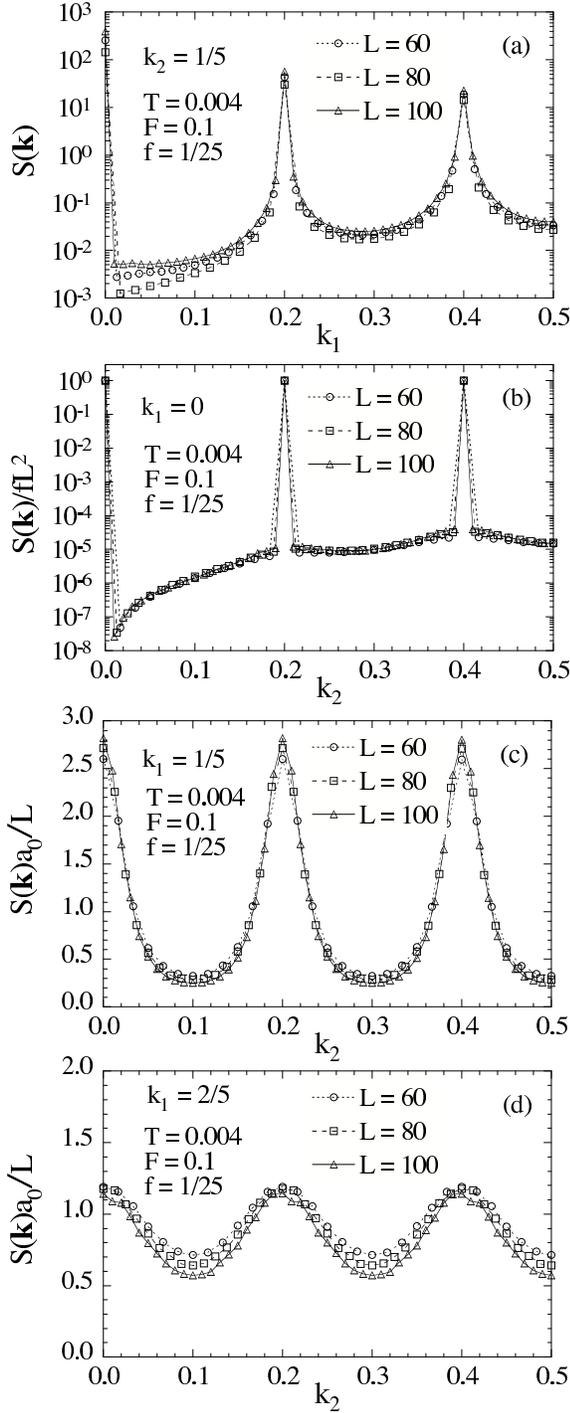}
\caption{Profiles of $S({\bf k})$ in various directions, for different system sizes $L$,
for the smectic phase at $F=0.10$, $T=0.004$. (a) $S({\bf k})$ vs. $k_1$ for fixed
$k_2=1/5$; (b) $S({\bf k})/fL^2$ vs. $k_2$ for fixed $k_1=0$;
(c) $S({\bf k})a_0/L$ vs. $k_2$ for fixed $k_1=1/5$; (d) $S({\bf k})a_0/L$
vs. $k_2$ for fixed $k_1=2/5$.  Note the logarithmic scale in (a) and (b).
}
\label{f7}
\end{figure}

To further illustrate the above results, we plot in Fig.\,\ref{f7} profiles of
S({\bf k}) along different paths through the first Brillouin zone, for various
$L\times L$ system sizes.  Fig.\,\ref{f7}a shows $S({\bf k})$ versus $k_1$
for fixed $k_2=1/5$.  The logarithmic vertical scale, varying over five orders
of magnitude, indicates how sharply the peaks are confined to the values
$k_1=1/5,2/5$; however $S({\bf k})$ appears to decrease continuously as
one moves away from the peak values.
Fig.\,\ref{f7}b shows the peaks indicating the smectic order,
i.e. $S({\bf k})/fL^2$ versus $k_2$ for fixed $k_1=0$.  We see that the peaks,
scaled by $N_c=fL^2$,  all have the same height for the different $L$, in
agreement with the scaling seen in Fig.\,\ref{f6}.  The logarithmic vertical
scale, dropping five orders of magnitude as one moves a single grid point
in $k$-space away from the peaks at $k_2=1/5,2/5$, shows that these
are indeed sharp Bragg peaks.  In Figs.\,\ref{f7}c,d we show the peaks
at finite $k_1$, plotting $S({\bf k})a_0/L$ versus $k_2$ for fixed
$k_1=1/5$ and $k_1=2/5$ respectively.  We see that these profiles,
when scaled by $1/L$, collapse reasonably well to a common curve
for the different sizes $L$, for all values of $k_2$.  
This is in agreement with the scaling found in Fig.\,\ref{f6}, and
suggests that $S({\bf k})$ is indeed $\delta$-function like in $k_1=1/5,2/5$
for all $k_2$.

\begin{figure}
\epsfxsize=7.5truecm
\epsfbox{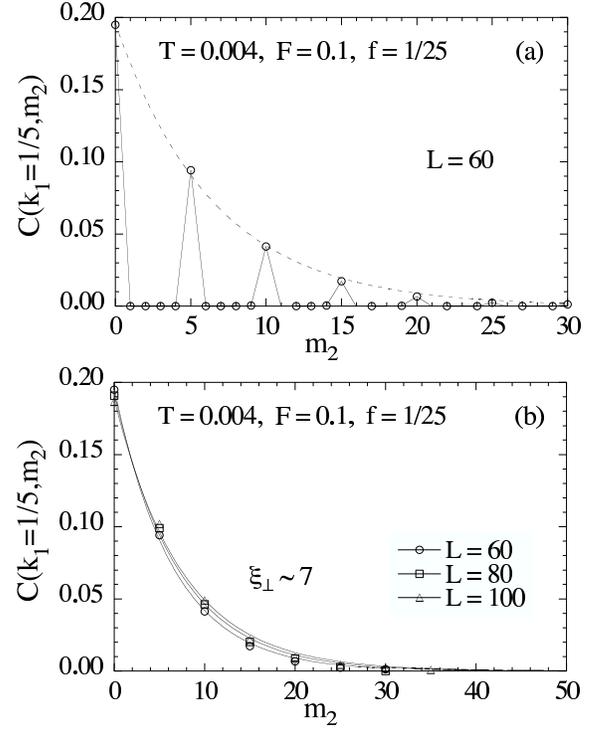}
\caption{Transverse correlation function $C(k_1=1/5,m_2)$ vs. $m_2$ at $F=0.1$, $T=0.004$.
(a) $C(k_1=1/5,m_2)$ for all integer values $m_2$ for the single size $L=60$; the
dashed line highlights the decaying envelop of the peaks, while the solid line interpolates 
between the data points.
(b) Peak values of $C(k_1=1/5,m_2)$ at values $m_2=5n$, $n$ integer, for different
sizes $L$;
solid lines are fits to the periodic exponential of Eq.\,(\ref{ePexp}).}
\label{f8}
\end{figure}

The finite widths in the $k_2$ direction of the peaks in 
$S({\bf k})$ at $k_1=1/5,2/5$, which do not narrow as
$L$ increases, indicate that the ordered smectic channels have only short range correlations
between them.  To see this explicitly,  consider the Fourier transform
of the charge density in each row of the grid at the wavevector corresponding to 
the periodic ordering within the smectic channels, 
i.e. ${\bf k}=(1/5){\bf b}_1$, and compute the correlations
of this Fourier amplitude between different rows.  This is given by the
correlation function $C(k_1=1/5,m_2)$, defined in Eq.\,(\ref{eCmixed}).
In Fig.\,\ref{f8}a we plot $C(k_1=1/5,m_2)$ versus $m_2$ for a $60\times 60$
system.  The correlation has peaks at values $m_2=5n$, $n$ integer,
and is essentially zero in between, indicating that the particles flow in periodically
spaced channels, and that the channels have a width of a single grid spacing.
The exponential decay of the peak heights indicates the short range
correlation between particles in different smectic channels.
In Fig.\,\ref{f8}b we plot only the peaks of $C(k_1=1/5,m_2)$, but for 
different system sizes $L\times L$.  The curves for different $L$ 
lie almost on top of each other and decay to zero, indicating a finite, size independent, 
correlation length $\xi_\perp$ transverse to the direction of the applied force ${\bf F}$.
To estimate $\xi_\perp$ we fit to a simple periodic exponential, 
\begin{equation}
C(k_1=1/5,m_2)\simeq A\left({\rm e}^{-m_2/\xi_\perp}+{\rm e}^{-(L-m_2)/\xi_\perp}\right)
\enspace,
\label{ePexp}
\end{equation}
and get values in the range $\xi_\perp \simeq 7.0\pm 0.5$ as $L$ varies from $60$ to $100$.
Thus correlations extend only slightly beyond nearest neighbor channels (which are separated
by $5$ grid spacings).

\begin{figure}
\epsfxsize=7.5truecm
\epsfbox{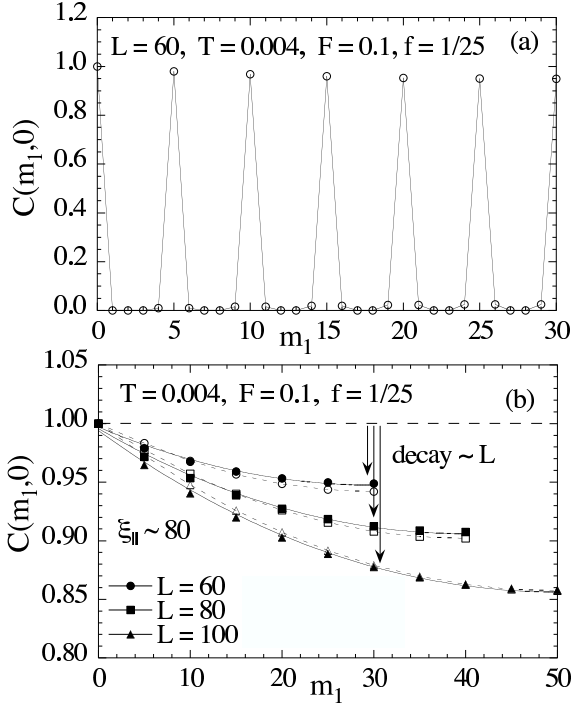}
\caption{Longitudinal correlation function $C(m_1,m_2=0)$ vs. $m_1$ at $F=0.1$, $T=0.004$.
(a) $C(m_1,0)$ for all integer values $m_1$ for the single size $L=60$.
(b) Peak values of $C(m_1, 0)$ at values $m_1=5n$, $n$ integer, for different
sizes $L$ (solid symbols); solid lines are fits to the periodic exponential of Eq.\,(\ref{ePexp2}).
Open symbols and dashed lines are fits to a one dimensional model (see text).
}
\label{f9}
\end{figure}

Next we compute the correlations within individual smectic channels.  In
Fig.\,\ref{f9}a we plot the real space correlation parallel to the driving
force, $C(m_1,m_2=0)$ versus $m_1$, for a system of size $60\times 60$.
Again we see sharply defined peaks at $m_1=5n$, $n$ integer, corresponding
to the periodic spacing of particles within the channel.  Moreover the
height of these peaks decays only slightly to a large finite value as $m_1\to L/2$,
as one would expect for long range order.  However, when we plot in 
Fig.\,\ref{f9}b the height of these peaks for different values of $L$ for 
system sizes $L\times L$, we now see behavior inconsistent with long range order.
The value of $C(m_1,0)$ at {\it any} given value of $m_1$ {\it decreases}
as $L$ increases; the magnitude of this decrease from unity is proportional to $L$.
Rather than indicating long range order, such behavior is consistent with a very
dilute but finite density of order destroying defects; when $L$ is small compared
to the average spacing between defects, then the probability to have a defect in the
system will be proportional to $L$, resulting in a decrease in the correlation
proportional to $L$.  

Indeed, since the smectic channels are essentially
decoupled from each other (as illustrated in Fig.\,\ref{f8}), each channel can
be thought of as an independent one dimensional system.  Although the charges in the 
channel have a bare long range logarithmic interaction, the uncorrelated motion of
charges in neighboring smectic channels will screen this log interaction, converting
it to an effective interaction that is finite ranged.  In equilibrium, such a one
dimensional system must be {\it disordered} at any finite $T$, and we expect that the
same will be true of a driven steady state.  To test this we perform
independent CTMC simulations of a one dimensional (1D) lattice gas of particles with average 
spacing $5$ and a nearest neighbor harmonic interaction with a spring constant $\kappa$.
Carrying out simulations for the same system sizes $L$ as in Fig.\,\ref{f9}b, we
adjust $\kappa$ to get the best fit to the correlations found in the original two
dimensional system.  This gives a reasonable value of $\kappa=0.0505$, and
our 1D results are shown as the open symbols and dashed lines in Fig.\,\ref{f9}b.
We see that the agreement is very good; the small deviations that exist are presumably 
due to the small but finite coupling between neighboring smectic channels that exists
in the original 2D model.  Having found $\kappa$, we can now simulate the 1D
model for much larger $L$, to see the exponential decay of correlations in the
model and to determine the correlation length $\xi_\|$.  We find for the 1D model,
$\xi_\|\simeq 86$.  For comparison, we can fit the data from the original 2D model
to a periodic exponential, 
\begin{equation}
C(m_1,m_2=0)\simeq c_1+c_2\left({\rm e}^{-m_1/\xi_\|}+{\rm e}^{-(L-m_1)/\xi_\|}\right)
\enspace,
\label{ePexp2}
\end{equation}
where $c_1=1/5$ is the average density of charges in a smectic channel, and
$c_2=4/5$ is chosen so that $C(0,0)=1$.  The resulting fits are shown as the
solid lines in Fig.\,\ref{f9}b, and give the values $\xi_\|\simeq 83, 80, 78$ for
sizes $L=60,80,100$, in good agreement with the 1D model.
We conclude that the smectic phase at high drive consists of weakly coupled
channels, characterized by a small transverse correlation length $\xi_\perp$. 
Within each channel particles have only finite range correlations,
but, for the case considered above, this longitudinal correlation length is comparable
to the size of the system, $\xi_\|\gtrsim L\gg \xi_\perp$, so that the particles
in a given channel appear to be ordered.

\begin{figure}
\epsfxsize=7.5truecm
\epsfbox{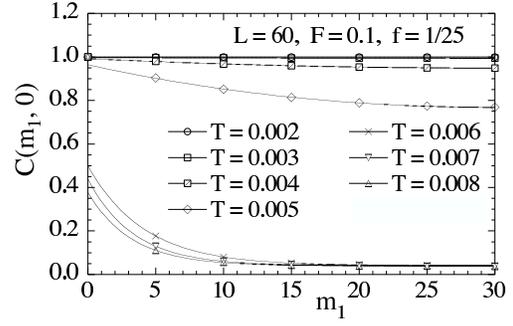}
\caption{Correlation function $C(m_1,m_2=0)$ vs. $m_1$ at $F=0.1$ and
various $T$, for a $60\times 60$ system.  
Only the peak values at $m_1=5n$, $n$ integer, are shown.
Solid lines are fits to an appropriate periodic exponential, as in Eq.\,(\ref{ePexp2}),
excluding the initial point at $m_1=0$ from the fit.}
\label{f10}
\end{figure}

\begin{figure}
\epsfxsize=8.5truecm
\epsfbox{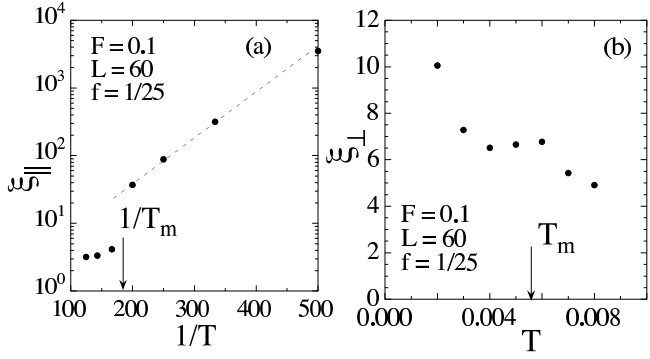}
\caption{(a) Longitudinal correlation length $\xi_\|$ vs. $1/T$, and (b) transverse
correlation length $\xi_\perp$ vs. $T$, at $F=0.1$ for a $60\times 60$ system.
}
\label{f11}
\end{figure}

We can next ask what happens if the system length parallel to the driving force ${\bf F}$
increases to be larger than the longitudinal correlation length, $L_1>\xi_\|$.
Increasing the system to size $L_1\times L_2=500\times 60$, so that this
condition is met, we found in Ref.\,[\onlinecite{Gotcheva}] 
that the smectic phase at $F=0.1$, $T=0.004$, becomes unstable to the liquid; for this system with bigger $L_1$, the smectic remains stable only at lower
$T$ such that the condition $L_1\lesssim\xi_\|$ is again obeyed.  To illustrate this
point further, we carry out simulations for a system of size $60\times 60$ at $F=0.1$,
but increasing $T$ so as to cross the melting line shown in the phase diagram of
Fig.\,\ref{f3}b.  In Fig.\,\ref{f10} we plot the resulting correlation functions
$C(m_1, m_2=0)$ versus $m_1$ for several different temperatures.  We see
clearly the transition from smectic to liquid at a temperature between $0.005$ 
and $0.006$.  To get an estimate of the longitudinal correlation length $\xi_\|$
we fit the data of Fig.\,\ref{f10} to a periodic exponential, as in Eq.\,(\ref{ePexp2}).
For the smectic, $T\le 0.005$, we set $c_1=1/5$ as appropriate for the
average density of charges in a smectic channel.  For the liquid, $T\ge 0.006$, we
set $c_1=1/25$, appropriate for the average density of charges in the liquid.
In both cases we find better results when we exclude the initial point at $m_1=0$,
$C(0,0)=1$, from the fit; we therefore keep $c_2$ as a free fitting parameter.
The resulting fits are shown as the solid lines in Fig.\,\ref{f10}.  The values of
$\xi_\|$ obtained from these fits are then plotted versus $1/T$ in Fig.\,\ref{f11}a.
The dashed straight line on the plot indicates an Ahrenius form, $\xi_\|
\sim {\rm e}^{T_0/T}$, for the divergence of
$\xi_\|$ in the smectic as $T\to 0$.  We see that
melting occurs when $\xi_\|\sim 30$, i.e. roughly half the system length.
We conclude that the smectic is only stable when $\xi_\|\gtrsim L/2$.  When the
correlation length becomes smaller than this, and one would expect
particles within the smectic channels to disorder, the entire smectic structure becomes
unstable to a liquid.  Since $\xi_\|$ diverges rapidly as $T\to 0$, however, one should expect to
see the smectic phase in any finite size system, at sufficiently low temperature.


Finally, we can also estimate the transverse correlation length $\xi_\perp$.
For the smectic we use an analysis of $C(k_1=1/5,m_2)$, similar to that
of Fig.\,\ref{f8}b, to determine $\xi_\perp$.  For the liquid, since there is
no periodic ordering, we use an analysis similar to that of Fig.\,\ref{f10}
applied to the real space correlation $C(x=0,y)=C(m_1=-m_2/2,m_2)$ 
versus $y=(\sqrt{3}/2)m_2$.  
Note that since
the direction $\hat a_2$ is not orthogonal to $\hat a_1$ (see Fig.\,\ref{f1}a)
it is necessary to use the argument $m_1=-m_2/2$ in order to measure a
strictly transverse correlation.
Our results for $\xi_\perp$ versus $T$ are shown in Fig.\,\ref{f11}b.
Unlike the rapid rise in $\xi_\|$ as $T$ decreases, 
we see only a small increase in $\xi_\perp$ as $T\to 0$.  In the liquid, 
$\xi_\perp$ and $\xi_\|$ are comparable.

\subsection{Smectic Phase - Low Drive}
\label{sLow}

\begin{figure}
\epsfxsize=7.5truecm
\epsfbox{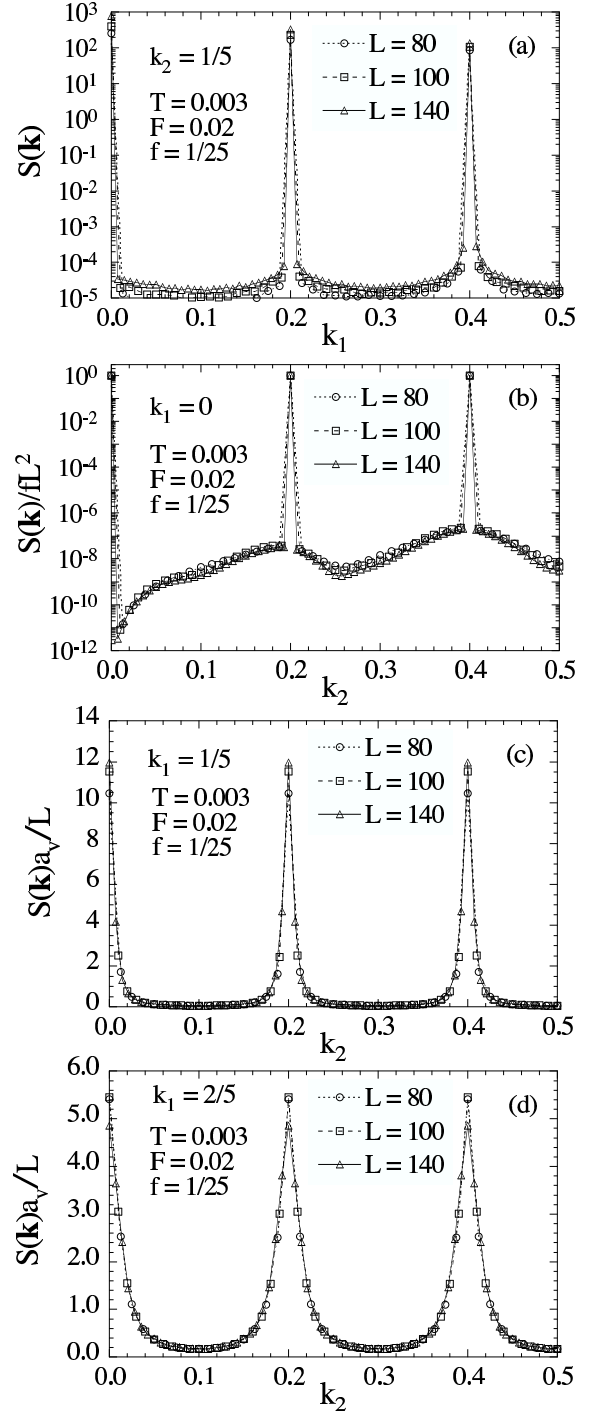}
\caption{Profiles of $S({\bf k})$ in various directions, for different system sizes $L$,
for the smectic phase at $F=0.02$, $T=0.003$. (a) $S({\bf k})$ vs. $k_1$ for fixed
$k_2=1/5$; (b) $S({\bf k})/fL^2$ vs. $k_2$ for fixed $k_1=0$;
(c) $S({\bf k})a_0/L$ vs. $k_2$ for fixed $k_1=1/5$; (d) $S({\bf k})a_0/L$
vs. $k_2$ for fixed $k_1=2/5$.  Note the logarithmic scale in (a) and (b).
}
\label{f12}
\end{figure}

We now consider the structure of the smectic in the limit of smaller
driving forces, in particular the case $F=0.02$, $T=0.003$, shown in Fig.\,\ref{f4}c.
In Fig.\,\ref{f12} we plot profiles of $S({\bf k})$ along different paths in the
first Brilloun zone, for different system sizes $L\times L$.  Comparing with the
analogous Fig.\,\ref{f7} for the high drive case, $F=0.1$, we see that the peaks
are now more sharply defined.  The peaks along the $k_2$ axis at $k_1=0$ in
Fig.\,\ref{f12}b continue to look like Bragg peaks, being only one grid point
wide and with heights scaling as $L^2$, thus indicating long range ordering
into smectic channels.
However the peak heights at finite $k_1=1/5,2/5$ in Figs.\,\ref{f12}c,d no longer
appear to scale $\sim L$ as do the corresponding peaks in Fig.\,\ref{f7}.

\begin{figure}
\epsfxsize=8.5truecm
\epsfbox{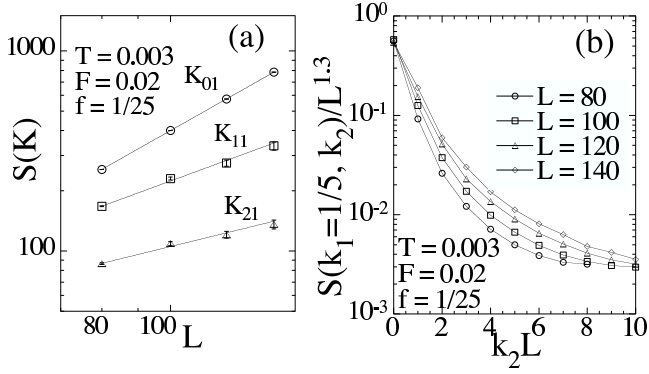}
\caption{Smectic phase at low drive, $F=0.02$, $T=0.003$.
(a) Scaling of peak heights $S({\bf K})$ vs. system size $L$. Straight lines indicate
good power law fits, $S({\bf K})\sim L^s$, with $s\simeq 2.0$ for ${\bf K}_{01}$,
$s\simeq 1.3$ for ${\bf K}_{11}$, and $s\simeq 0.87$ for ${\bf K}_{21}$.
(b) Attempted scaling collapse of $S(k_1=1/5,k_2)/L^{1.3}$ vs. $k_2L$.
}
\label{f13}
\end{figure}

In Fig.\,\ref{f13}a we plot the heights of various peaks $S({\bf K})$ versus
$L$ for various system sizes $L\times L$.  While the smectic peak $S({\bf K}_{01})$
scales $\sim L^2$ as expected, we find $S({\bf K}_{11})\sim L^{1.3}$,
more divergent than the $\sim L$ behavior found at higher drive.  This suggests the
possibility of longer range, perhaps algebraic, correlations between the different
smectic channels.  For algebraically diverging peaks, however, we expect that not
only the peak height must scale, but the entire peak profile should scale.  The expected
scaling relation is \cite{Moon},
\begin{equation}
S(k_1=1/5,k_2)\sim L^{1.3}f(k_2L)
\label{eSscale}
\end{equation}
where $f(x)$ is a scaling function.  In Fig.\,\ref{f13}b we test this scaling prediction
by plotting $S(k_1=1/5,k_2)/L^{1.3}$ versus $k_2L$ for different sizes $L$.
We clearly do {\it not} find the scaling collapse expected for algebraic correlations.

\begin{figure}
\epsfxsize=7.5truecm
\epsfbox{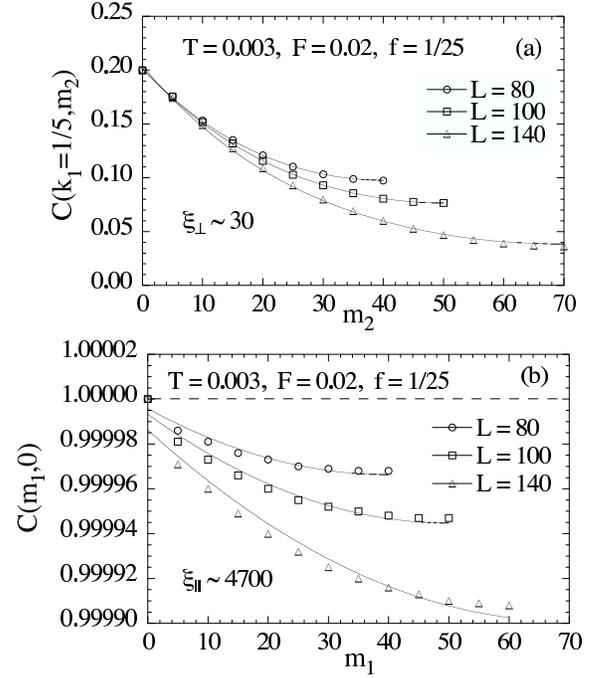}
\caption{Smectic phase at low drive, $F=0.02$, $T=0.003$, for various
system sizes $L\times L$.
(a) Transverse correlation function $C(k_1=1/5,m_2)$ vs. $m_2$, and
(b) longitudinal correlation function $C(m_1,m_2=0)$ vs. $m_1$.
Solid lines are fits to periodic exponentials as in Eqs.\,(\ref{ePexp}) and
(\ref{ePexp2}) and determine the correlation lengths $\xi_\perp$ and
$\xi_\|$.
}
\label{f14}
\end{figure}

To explain the above behavior, we consider the transverse correlation
function $C(k_1=1/5,m_2)$, which we plot versus $m_2$ for different
system sizes $L\times L$ in Fig.\,\ref{f14}a.  The solid lines are fits to
the periodic exponential of Eq.\,(\ref{ePexp}), and give the values
$\xi_\perp\simeq 29.6, 30.8, 29.7$ for sizes $L=80, 100,140$, respectively.
Our results thus consistently indicate short range order between the smectic
channels, with a finite transverse correlation length $\xi_\perp\simeq 30$.
The absence of the expected $\sim L$ scaling in Figs.\,\ref{f12}c,d is then a finite
size effect due to the correlation length $\xi_\perp$ being comparable
to the system length $L$.

We similarly estimate the longitudinal correlation length by plotting
$C(m_1,m_2=0)$ versus $m_1$, for different system sizes $L\times L$,
in Fig.\,\ref{f14}b.  Fitting to the periodic exponential of Eq.\,(\ref{ePexp2}),
with $c_1=1/5$ and $c_2=4/5$, we find $\xi_\|\sim 4700$.  We conclude that
the smectic phase at small drive is qualitatively the same as that at large drive,
except for having larger, but still finite, correlation lengths.

\begin{figure}
\epsfxsize=7.5truecm
\epsfbox{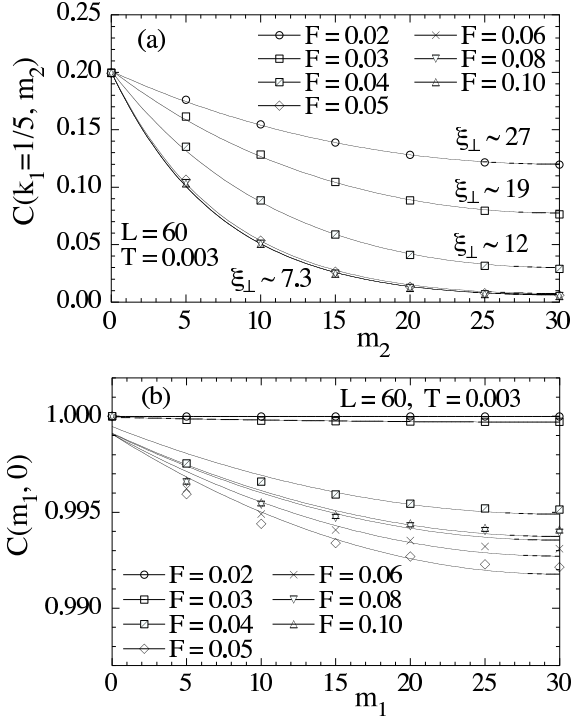}
\caption{Smectic phase at $T=0.003$, for various driving forces $F$
in a $60\times 60$ size system.
(a) Transverse correlation function $C(k_1=1/5,m_2)$ vs. $m_2$, and
(b) longitudinal correlation function $C(m_1,m_2=0)$ vs. $m_1$.
Solid lines are fits to periodic exponentials as in Eqs.\,(\ref{ePexp}) and
(\ref{ePexp2}) and determine the correlation lengths $\xi_\perp$ and
$\xi_\|$.
}
\label{f15}
\end{figure}

\begin{figure}
\epsfxsize=8.5truecm
\epsfbox{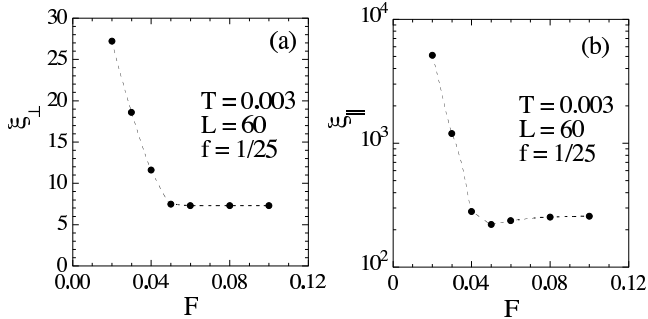}
\caption{(a) Transverse correlation length $\xi_\|$ and (b) longitudinal
correlation length $\xi_\perp$ vs. $F$, at $T=0.003$ for a $60\times 60$ system.
}
\label{f16}
\end{figure}

Finally we consider behavior as the driving force $F$ varies.
In Fig.\,\ref{f15}a we plot the transverse correlation function
$C(k_1=1/5,m_2)$ versus $m_2$, for $T=0.003$ in a $60\times 60$
system, for various values of $F$ from the low drive case considered
above, $F=0.02$, to the high drive case considered previously, $F=0.1$.
Solid lines are fits to the periodic exponential of Eq.\,(\ref{ePexp}).  In Fig.\,\ref{f15}b
we plot the corresponding longitudinal correlation function $C(m_1,m_2=0)$ 
versus $m_1$.  Solid lines are fits to the periodic exponential of Eq.\,(\ref{ePexp2}),
with $c_1=1/5$ and $c_2=4/5$.  From these fits we estimate the transverse and
longitudinal correlation lengths, $\xi_\perp$ and $\xi_\|$, which are plotted
in Figs.\,\ref{f16}a,b.
We see that the transverse correlation length $\xi_\perp$ grows as 
$F$ decreases below $0.05$.  For 
$F\ge 0.05$, $\xi_\perp$ levels off to a constant, $\xi_\perp\sim 7$.
In fact, for $F\ge 0.05$, the transverse correlation $C(k_1=1/5,m_2)$
shown in Fig.\,\ref{f15}a is completely independent of $F$.  The longitudinal
correlation length $\xi_\|$ grows exponentially as $F$ decreases below $0.04$,
has a shallow minimum at $F=0.05$ (presumably related to the minimum in the
melting line $T_{\rm m}(F)$ that occurs nearby) and then saturates to a constant above
$F=0.06$.  The longitudinal correlation $C(m_1,m_2=0)$ shown in 
Fig.\,\ref{f15}b is independent of $F$ for $F>0.06$.   These results strongly
suggest that no new behavior will be seen by increasing $F$ to even larger values, and
we have verified this explicitly by simulating up to $F=2.0$.

The above cross over point between low and high drives, and its value
$F_{\rm cr}\simeq 0.05$, can be understood from Fig.\,\ref{f2}b.  Consider the system in its
$F=0$ ground state.  The interaction energy to move
a given charge forward parallel to ${\bf F}$ is $\Delta{\cal H}_1\simeq 0.0627$;
the rate to make this move is $W=W_0{\rm e}^{-(\Delta{\cal H}_1-F)/2T}$.
Once this charge has moved forward, the interaction energy to move the
neighboring charge in the same row forward is $\Delta{\cal H}_2\simeq 0.0270$;
the rate to make this move is $W_{\rm f}=W_0{\rm e}^{-(\Delta{\cal H}_2-F)/2T}$.
This needs to be compared against the rate for the first charge to move back to
its original position, $W_{\rm b}=W_0{\rm e}^{(\Delta{\cal H}_1-F)/2T}$.
The ratio of these last two rates is,
\begin{equation}
{W_{\rm f}\over W_{\rm b}} = {\rm e}^{(2F-\Delta{\cal H}_1-\Delta{\cal H}_2)/2T}
\enspace,
\label{eRateRatio}
\end{equation}
and so the two rates are equal when $F_{\rm cr}=(\Delta{\cal H}_1-\Delta{\cal H}_2)/2=0.045$,
which agrees with our observations in Figs.\,\ref{f15} and \ref{f16}.
When $F>F_{\rm cr}$, the probability that the neighboring charge  moves
forward along with the first charge is larger than the probability that the first charge falls
back into place.  Once the second charge moves forward, the remaining charges in the
row will follow suite.  Therefore when $F\gg F_{\rm cr}$, 
virtually {\it all} moves are those which advance a charge in the direction of 
the applied force ${\bf F}$; charges in the smectic channels move continuously 
forward row by row.  For $F<F_{\rm cr}$, charges which advance forward parallel to 
${\bf F}$ will more often than not fall backwards to their original position on the next move.  
The system spends finite time with no net motion, in between randomly occurring avalanches
that advance an entire row of charges forward.  The result is a stick-slip type of
motion.

\begin{figure}
\epsfxsize=9.0truecm
\epsfbox{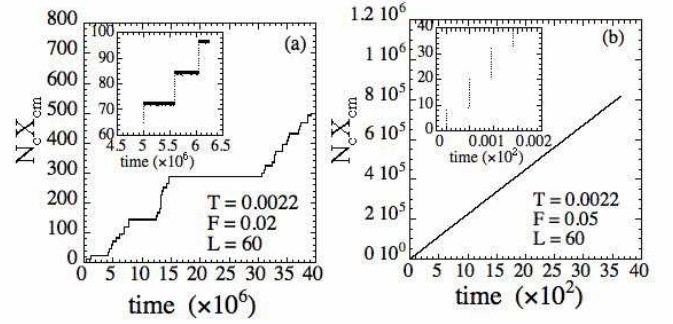}
\caption{Center of mass displacement parallel to the driving force for
$T=0.0022$, $L=60$. $N_cX_{\rm cm}$ vs. simulation clock
time $t$ for (a) $F=0.02$ and (b) $F=0.05$.  Insets show an
expanded picture on a short time scale.
}
\label{f17}
\end{figure}

The above scenario is illustrated in Fig.\,\ref{f17} where we plot
the center of mass displacement parallel to ${\bf F}$ times the number of 
charges, $N_cX_{\rm cm}$, versus the simulation clock tme $t$.
We show results for $T=0.0022$, slightly lower than the temperature $0.003$
considered above, for a system of size $60\times 60$.   Fig.\,\ref{f17}a,
for $F=0.02<F_{\rm cr}$, shows the step like advancement forward of the system,
characteristic of stick-slip motion.  The inset shows an expanded scale for
short time.  The height of each step is exactly $12$ grid spacings, corresponding
to the advance of all $12$ charges in a given smectic channel.  Along the plateau
of each step we see motion one grid space forward, followed by one grid space
backwards, with no net motion.
In Fig.\,\ref{f17}b we show results for $F=0.05\gtrsim F_{\rm cr}$.
We see that motion is perfectly linear in time.  The inset shows that
the system moves smoothly forward, with one row advancing immediately 
after another.

\subsection{Liquid Phase}

\begin{figure}
\epsfxsize=8.5truecm
\epsfbox{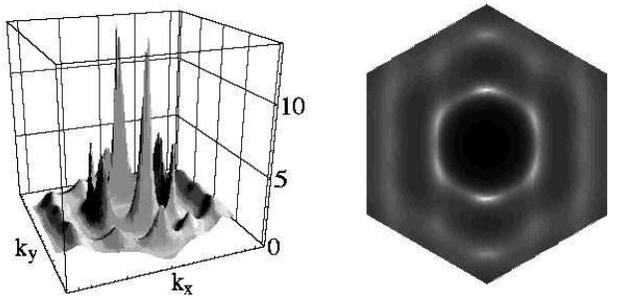}
\caption{Structure function $S({\bf k})$ for system of size $120\times 120$
in the liquid state at $T=0.003$ and $F=0.05$.
}
\label{f18}
\end{figure}

We now briefly consider the liquid phase.  The liquid phase shown in
Fig.\,\ref{f4}b, at $F=0.02$, $T=0.008$, appears fairly structureless.
However as $T$ decreases, and the finite correlation lengths grow,
local structure develops.  As discussed in section \ref{sHigh} the
melting line $T_{\rm m}(F)$ decreases as the size of the system
increases.  Although, for $L=60$, $T_{\rm m}(F)$ always lies above
$T=0.003$ (see Fig.\,\ref{f3}b), when $L$ increases, the minimum
of $T_{\rm m}(F)$ near $F\sim 0.03$ dips below $0.003$.  In
Fig.\,\ref{f18} we plot the structure function $S({\bf k})$ for
a system of size $120\times 120$ at $T=0.003$ and driving
force $F=0.05$.  Although one sees prominent peaks at the reciprocal 
lattice vectors corresponding to  the $F=0$ ground state, the system is in
a liquid state with short range translational order.  The heights of the
peaks are small compared to those in a more ordered (i.e. smectic or solid)
state, and as $L$ increases, the peak heights stay finite rather than diverging with 
system size.

%

\begin{figure}
\epsfxsize=8.5truecm
\epsfbox{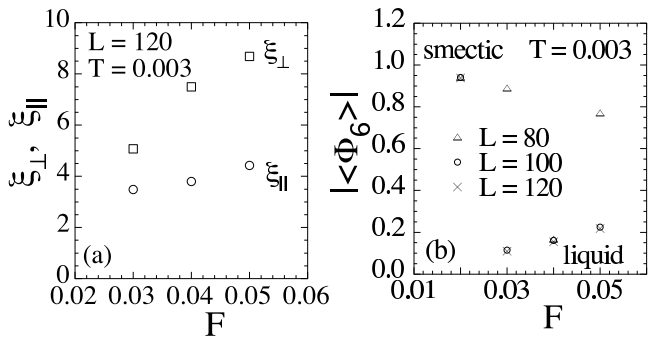}
\caption{(a) Correlation lengths $\xi_\perp$ and $\xi_\|$ vs. $F$ for a system of size
$120\times 120$ in the liquid state at $T=0.003$.
(b) Orientational order parameter $|\langle\Phi_6\rangle|$ 
vs. $F$ for several systems of size $L\times L$
at $T=0.003$.  The larger valued data points are for the smectic phase; the lower data
points are for the liquid phase.
}
\label{f19}
\end{figure}

Next we consider how behavior in the liquid varies with the driving force $F$.
In Fig.\,\ref{f19}a we plot the longitudinal and transverse
correlation lengths $\xi_\perp$ and $\xi_\|$ versus $F$.  We obtain
$\xi_\|$ and $\xi_\perp$ 
by fitting to the correlation functions $C(m_1,m_2=0)$ and $C(x=0,y)$
in the same way as we have done earlier in constructing Fig.\,\ref{f11}.  We see that
$\xi_{\perp}$ and $\xi_\|$ increase with increasing $F$, and that $\xi_\perp\sim 2\xi_\|$.
That order extends further in the transverse than the longitudinal direction can
be seen by noting that the transverse peaks in $S({\bf k})$, shown in Fig.\,\ref{f18},
are larger than the peaks with a longitudinal component.  The same observation
was made in our earlier work \cite{Gotcheva} for a much larger system
at higher driving force.

We also investigate orientational order in the liqud.  In Fig.\,\ref{f19}b we plot 
the absolute value of the $6$-fold orientational order parameter $|\langle\Phi_6\rangle|$ versus $F$
at $T=0.003$,  for several different system sizes $L\times L$.  Depending on
the value of $L$ and the corresponding value of $T_{\rm m}(F;L)$, 
the system is either in the smectic state (with a high value of
$|\langle\Phi_6\rangle|$) or in the liquid state (with a low value of $|\langle\Phi_6\rangle|$).
Even in the liquid $|\langle\Phi_6\rangle|$ is finite because of the $6$-fold rotational
symmetry of the underlying triangular grid.  We see that, as $F$ increases in the
liquid, $|\langle\Phi_6\rangle|$, like $\xi_\perp$ and $\xi_\|$, increases.  Thus
both translational and orientational order in the liquid increase as the driving
force $F$ increases.

\subsection{Dynamics}

\begin{figure}
\epsfxsize=8.5truecm
\epsfbox{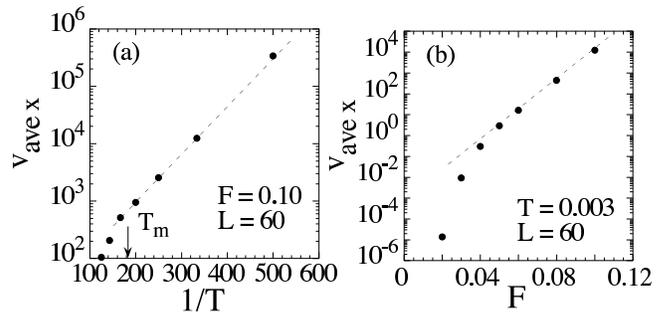}
\caption{Average velocity $v_{{\rm ave}\, x}$ parallel to the driving force ${\bf F}$
for a $60\times 60$ system.  (a) $v_{{\rm ave}\, x}$ vs. $1/T$ for fixed $F=0.1$ in the
high drive limit. (b) $v_{{\rm ave}\, x}$ vs. $F$ for fixed $T=0.003$ in the smectic.
The dashed lines indicate the exponential dependence of $v_{{\rm ave}\, x}$ on $F$ and
$1/T$.
}
\label{f20}
\end{figure}

The preceding sections have dealt with the structural behavior of the driven system. 
In this section we consider some of the dynamical behavior.  In Fig.\,\ref{f20}
we plot results for the average velocity $v_{{\rm ave}\,x}$ parallel to the driving
force ${\bf F}=F\hat x$, for a $60\times 60$ size system.  
In Fig.\,\ref{f20}a we show results for fixed $F=0.10>F_{\rm cr}$, in the high 
drive limit, versus $1/T$.  The dashed line for $1/T_{\rm m}<1/T$ shows
the exponential dependence on $1/T$ in the smectic phase, $v_{{\rm ave}\,x}\sim {\rm e}^
{T_0/T}$.  Fitting to this form we find $T_0\simeq 0.02$.  We can understand this
value as follows.  As discussed in section \ref{sLow}, at such high $F$ virtually
all moves in CTMC result in the advance of a charge forward; the charges in
the smectic channels move steadily forward one channel at a time.  The average
velocity is set by the rate for the charges in a given channel to move forward, which
in turn is set by the rate for the first charge in the channel to move forward (all other
charges in that channel moving forward on a much more rapid time scale).  The rate
for the first charge in a channel to move forward is 
$\sim {\rm e}^{-\Delta U_{{\rm min}\,1}/2T}$,
where $-\Delta U_{{\rm min}\,1}=F-\Delta{\cal H}_1$, with $\Delta{\cal H}_1\simeq 0.06$
from Fig.\,\ref{f2}b.  Thus we have $T_0=(F-\Delta{\cal H}_1)/2=0.02$.

In Fig.\,\ref{f20}b we show results for $v_{{\rm ave}\,x}$ vs. $F$, for fixed $T=0.003<T_{\rm m}$, in the smectic phase.  The dashed line shows the exponential dependence on $F$
in the high drive limit, $F>F_{\rm cr}\simeq 0.045$, $v_{{\rm ave}\,x}\sim
{\rm e}^{F/F_0}$.  From the preceding discussion, we expect $F_0=2T=0.006$,
and we find this value gives an excellent fit to the data.  

%

We have also considered the dependence of the average velocity on the system size.
In Table\,\ref{t1} we list the values of $v_{{\rm ave}\,x}$ for various system sizes, with different
aspect ratios $L_1/L_2$, at $F=0.10$, $T=0.004$, in the high drive smectic.  In agreement
the discussion at the end of section \ref{sLowT} we see that $v_{{\rm ave}\,x}$
scales roughly proportional to the length of the system $L_1$ parallel to the
driving force.

\begin{table}[htdp]
\caption{Average velocity $v_{{\rm ave}\,x}$ for various system sizes on a triangular grid
at $F=0.10$, $T=0.004$, in the high drive smectic.}
\begin{ruledtabular}
\begin{tabular}{|c|c|c|c|c|c|c|}
$L_1$ &60 &60 & 120 & 120 & 120 & 240  \\
$L_2$ &30 & 60 & 30 & 60 & 120 & 60 \\
$v_{{\rm ave}\,x}$ & 2569 & 2550& 4971 & 4930& 4978 & 7135 \\
\end{tabular}
\end{ruledtabular}
\label{t1}
\end{table}

\begin{figure}
\epsfxsize=8.5truecm
\epsfbox{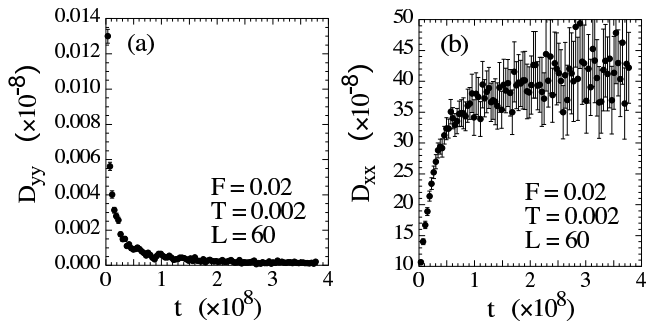}
\caption{Center of mass diffusion constants (a) $D_{yy}$ and (b) $D_{xx}$ vs. time $t$ for
the smectic phase in the low drive limit, $F=0.02$, $T=0.002$, for a system of
size $60\times 60$. That $D_{yy}\to 0$ indicates the system is transversely pinned.
}
\label{f21}
\end{figure}

\begin{figure}
\epsfxsize=8.5truecm
\epsfbox{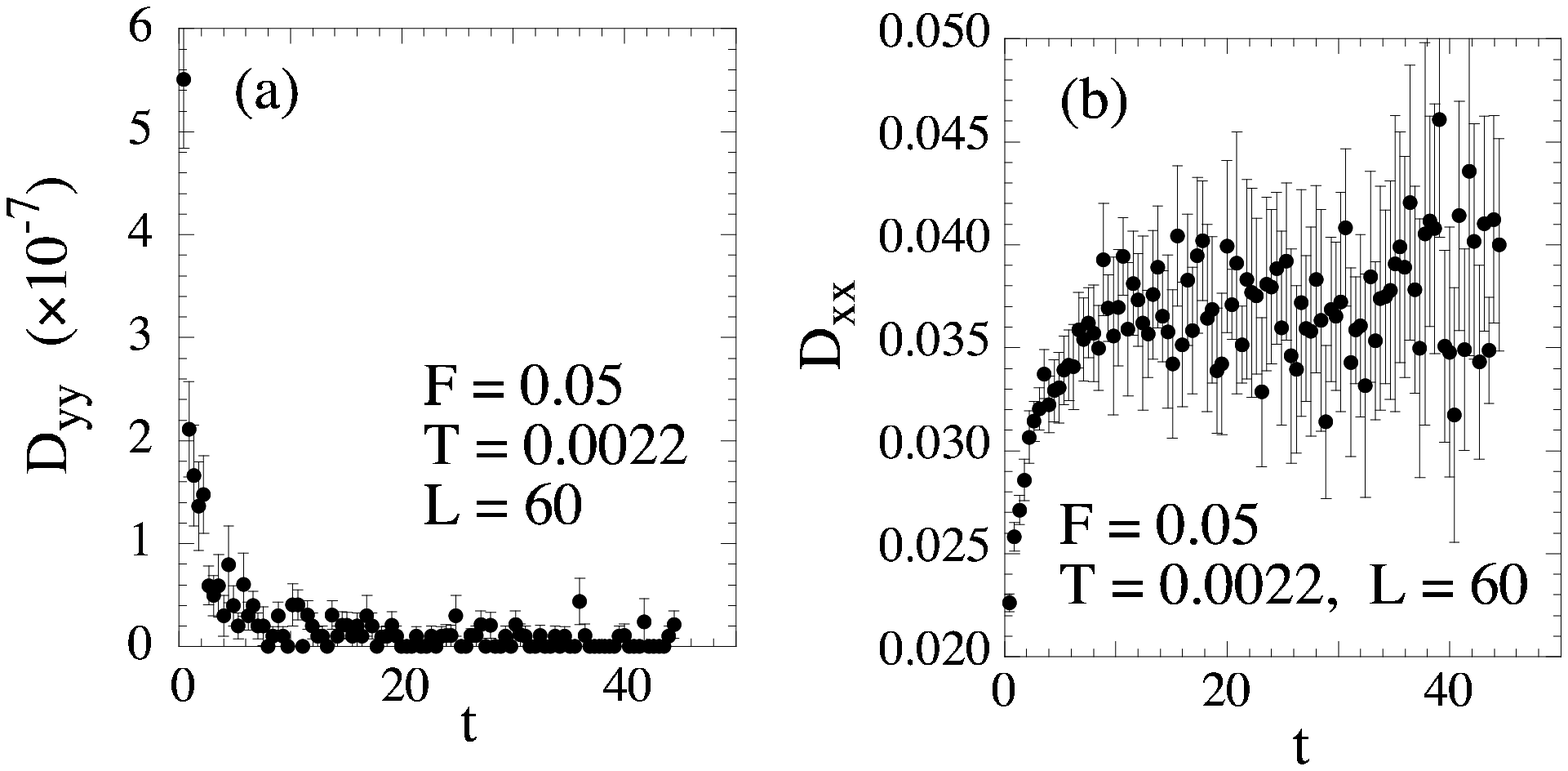}
\caption{Center of mass diffusion constants (a) $D_{yy}$ and (b) $D_{xx}$ vs. time $t$ for
the smectic phase in the high drive limit, $F=0.05$, $T=0.0022$, for a system of
size $60\times 60$. That $D_{yy}\to 0$ indicates the system is transversely pinned.
}
\label{f22}
\end{figure}

\begin{figure}
\epsfxsize=8.5truecm
\epsfbox{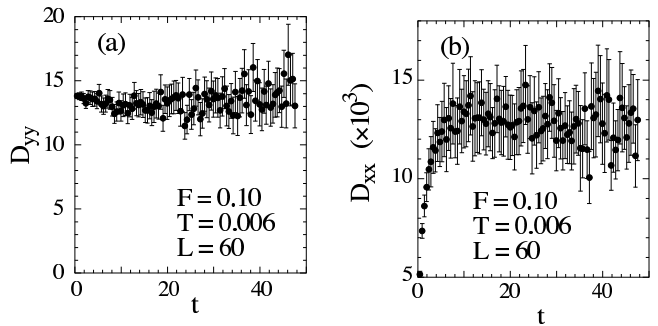}
\caption{Center of mass diffusion constants (a) $D_{yy}$ and (b) $D_{xx}$ vs. time $t$ for
the liquid phase at, $F=0.10$, $T=0.006$, for a system of
size $60\times 60$.
}
\label{f23}
\end{figure}

Next we consider the diffusion of the center of mass about its average motion.
To compute ${\bf D}(t)$ we need to compute the correlation between states
of the system separated by time $t$.  Since we are interested in the long $t$
limit, computing ${\bf D}(t)$ accurately thus requires much longer simulations
than were needed to compute the structural (equal time) correlations.  We therefore
present results only for several typical cases.
In Fig.\,\ref{f21} we plot $D_{yy}(t)$ and $D_{xx}(t)$, defined by Eq.\,(\ref{eDtensor2}),
versus the simulation clock time $t$, for the smectic 
in the low drive limit of $F=0.02$, $T=0.002$,
for a $60\times 60$ size system. We see that $D_{yy}$ decays to zero, indicating
that the system is transversely pinned.  $D_{xx}$ saturates to a finite constant
as $t$ increases, indicating a random walk motion about the average center of mass
position.  In Fig.\,\ref{f22} we similarly plot $D_{yy}$ and $D_{xx}$ for the smectic
in the high drive limit of $F=0.05$, $T=0.0022$.  Again we see that $D_{yy}\to 0$ and
the system is transversely pinned, while $D_{xx}$ saturates to a finite constant.
In Fig.\,\ref{f23} we plot $D_{yy}$ and $D_{xx}$ for the liquid at $F=0.10$, $T=0.006$,
just above the melting transition.  In this case both $D_{yy}$ and $D_{xx}$ approach
finite constants as $t$ increases; as expected, the liquid is not transversely pinned.


Since both $D_{xx}$ and $v_{{\rm ave}\,x}$ approach constants in the long time
limit, a convenient measure of the strength of fluctuations about the average motion
is given by $D_{xx}/v_{{\rm ave}\,x}$.  For the low drive smectic of Fig.\,\ref{f21}
we find $D_{xx}/v_{{\rm ave}\,x}\simeq 40$.  This is consistent with our interpretation
of this region as being one of stick-slip motion.  In this case we expect that the motion
of rows of charges forward will constitute a Poisson process with avalanches occurring
at a rate $\lambda$.  At each avalanche $n_r\sqrt{f}L$ charges move forward, where
$\sqrt{f}L$ is the number of charges in a given smectic channel, and $n_r$ is the
number of correlated channels.  The average center of mass displacement  in time $t$
is then $\Delta X_{\rm cm}=(n_r\sqrt{f}L/N_c)\lambda t =(n_r/[\sqrt{f}L])\lambda t$, 
where we used $N_c=fL^2$ is the total number of charges.
Because it is a Poisson process, the variance of the number of avalanches
is equal to the average, and so 
the fluctuation about this average displacement is $(\Delta X_{\rm cm})^2
=(n_r/[\sqrt{f}L])^2\lambda t$.  This yields the ratio $D_{xx}/v_{{\rm ave}\,x}=
N_c(\Delta X_{\rm cm})^2/2\Delta X_{\rm cm} = n_r\sqrt{f}L/2$.  For
$f=1/25$ and $L=60$ we get $D_{xx}/v_{{\rm ave}\,x}=6n_r$.  At low $T$ and
$F$ the correlation length $\xi_\perp$ can get large (see Fig.\,\ref{f16}); if several
channels are correlated, the ratio $40$ can be attained.

For the high drive case of Fig.\,\ref{f22}, we find the ratio
$D_{xx}/v_{{\rm ave}\,x}\simeq 0.00025$.  In this limit where
rows of channels move steadily forward one after the other, longitudinal fluctuations
are greatly suppressed.   Finally, for the liquid case of Fig.\,\ref{f23},
we find $D_{xx}/v_{{\rm ave}\,x}\simeq 25$.  In the liquid, the system
is structurally disordered and the motion of the charges is largely uncorrelated.
Fluctuations about the center of mass motion are correspondingly enhanced.

\section{Results on a square grid}
\label{sSquareResults}

\begin{figure}
\epsfxsize=5.5truecm
\epsfbox{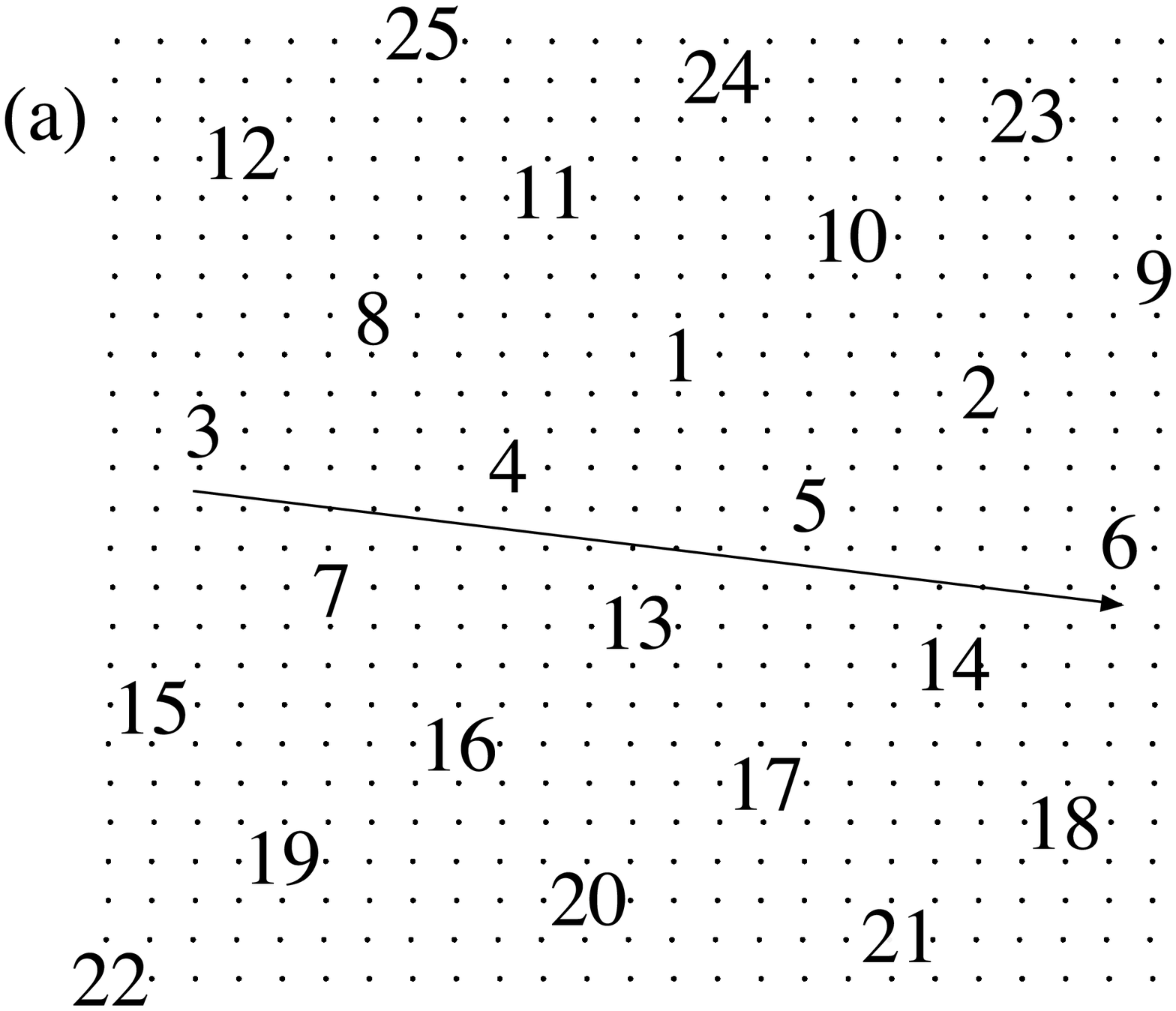}
\vskip .5truecm
\epsfxsize=8.0truecm
\epsfbox{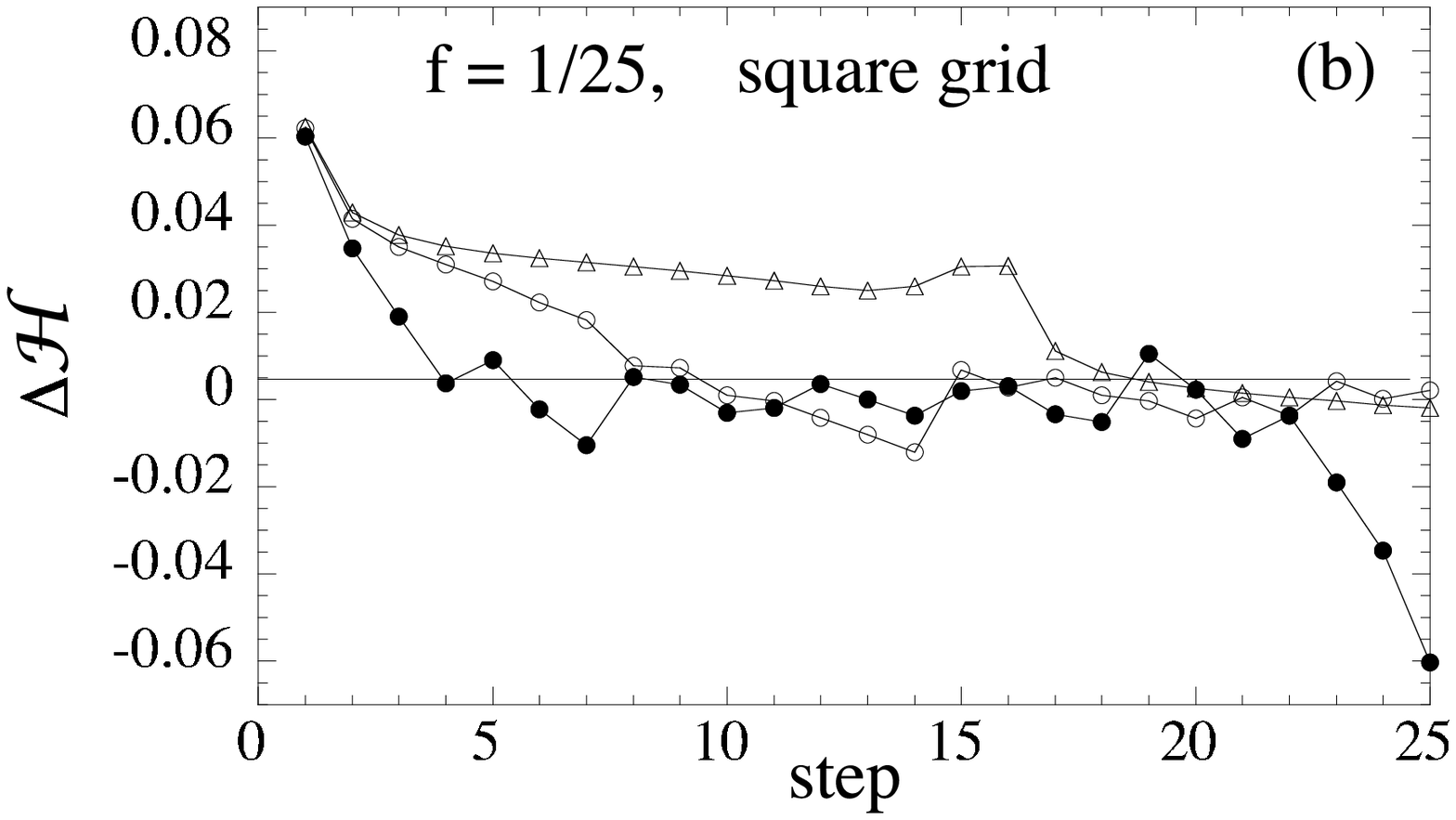}
\caption{CTMC on a square grid with charge density $f=1/25$
at $T\to 0$ and $F>F_c$, with ${\bf F}$ parallel to the $\hat a_1$ axis.
(a) Ground state charge lattice for a $25\times 25$ square grid. 
Numbers denote the locations of the charges in the
ground state.  The value of each number indicates the step on which that charge
moves forward.   (b) The change in interaction
energy $\Delta{\cal H}$ at each step as charges move forward. ($\bullet$) are for
a $25\times 25$ grid and correspond to the moves in (a); ($\circ$) and ({\scriptsize $\triangle$})
are the beginnings of similar sequences for $50\times 50$ and $100\times 100$ grids.
}
\label{f24}
\end{figure}

We now consider the behavior of the driven Coulomb gas on a periodic
square grid of sites.  We consider only CTMC dynamics for the same charge
density of $f=1/25$ that was considered above for the triangular grid.
We first consider behavior in the limit $T\to 0$.  In Fig.\,\ref{f24}a
we show the equilibrium ground state configuration for $F=0$.  The charges 
occupy the sites of a $5\times 5$ square sub lattice of the grid.  The basis
vectors of this sub lattice, ${\bf c}_1=3\hat x-4\hat y$ and 
${\bf c}_2=4\hat x+3\hat y$, are clearly {\it not} aligned with the grid basis 
vectors $\hat a_1=\hat x$ and $\hat a_2=\hat y$, nor with the driving force 
${\bf F}=F\hat x$ that we apply.  This will produce some interesting effects.

When $F>F_c\simeq 0.06$ the charges will start to move forward parallel to ${\bf F}$,
according to the order in which they most lower the system energy.  In Fig.\,\ref{f24}a
we number the charges in the order in which they move in a particular CTMC pass, 
and in Fig.\,\ref{f24}b
we give the change in interaction energy $\Delta{\cal H}$ associated with each move,
as was done for the triangular grid in Fig.\,\ref{f2}.  Charges move forward in the $\hat x$
direction in an order dictated by their position along 
the ${\bf c}={\bf c}_1+{\bf c}_2$ direction,
as indicated by the arrow drawn in Fig.\,\ref{f24}a.   If one follows a path along the
direction of ${\bf c}$, using periodic boundary conditions, one finds that the path
closes upon itself only after one has passed through all the charges in the ground state.
Thus there is no row by row motion as there was for the case of the triangular grid,
and hence no oscillation in $\Delta{\cal H}$ as a function of simulation step.

\subsection{High Drive}
\label{sSqHD}

\begin{figure}
\epsfxsize=8.0truecm
\epsfbox{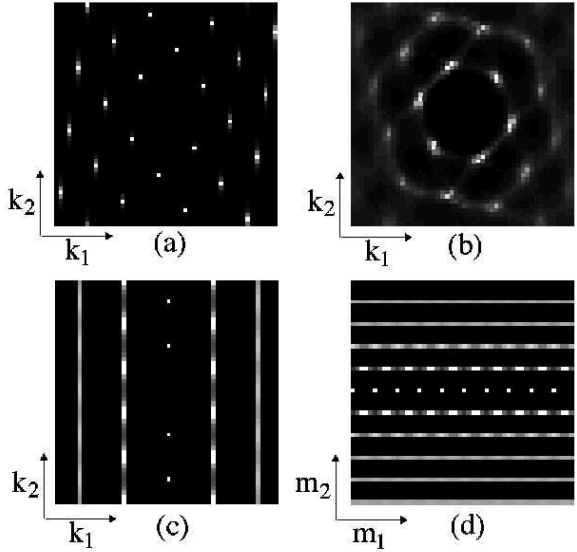}
\caption{(a-c) Intensity plot of structure function
$S({\bf k})$ for a $50\times 50$ size system at $T=0.004$, ${\bf F}=0.1\hat a_1$, 
starting from the ground state of Fig.\,\protect\ref{f24}a.  (a) $S({\bf k})$
averaged over $3750$ passes, after an initial $2500$ passes of equilibration; 
(b) $S({\bf k})$ averaged over $5\times 10^6$ passes, after discarding an initial 
$5\times 10^6$ passes;
(c) $S({\bf k})$ averaged over $2.5\times 10^7$ passes, after discarding an initial
$3.75\times 10^7$ passes. (d) Intensity plot of real space correlations 
$C(m_1,m_2)$ corresponding to the smectic $S({\bf k})$ of (c).
}
\label{f25}
\end{figure}

We now consider behavior at finite temperature and high drive.  We
simulate a $50\times 50$ size system, starting from the $F=0$ ground state,
at the values $T=0.004$, $F=0.10$.  In Fig.\,\ref{f25}a we show an intensity
plot of the structure function $S({\bf k})$ at the initial stage of the simulation;
our results are averaged over $1250$ CTMC passes after an initial $2500$ passes were 
discarded for equilibration.  We see peaks at wavevectors ${\bf K}$ corresponding
to the ordered $F=0$ ground state.  The peaks remain sharp in the $k_1$ direction, but are
somewhat smeared out in the transverse $k_2$ direction, suggesting a moving
lattice with anisotropic translational correlations.  If we simulate longer however,
this moving ground state lattice undergoes a change of structure.  In Fig.\,\ref{f25}b 
we show $S({\bf k})$ averaged over $5\times 10^6$ passes, after discarding an initial 
$5\times 10^6$ passes.
We see clearly a $6$-fold orientational order in the position of the peaks, which
are aligned with one of the diagonals of the square grid.  Fig.\,\ref{f25}b
is reminiscent of the floating triangular lattice (algebraic translational order) that is 
seen in {\it equilibrium} simulation \cite{Franz} of more dilute systems on a square grid,
however without a finite size scaling analysis we cannot be certain of the
nature of translational order in the system.
Finally, however, if we simulate even longer, the structure changes yet again to
a smectic phase with channels oriented parallel to ${\bf F}$.  In Fig.\,\ref{f25}c
we show $S({\bf k})$ averaged over $2.5\times 10^7$ passes, after discarding an initial
$3.75\times 10^7$ passes.  We see clearly the same smectic structure that we
saw for the triangular grid in Fig.\,\ref{f4}d.  The extremely long (compared to the
triangular grid) time it takes for the system to order into the smectic results because
the initial ground state configuration is ordered with a set of reciprocal lattice
vectors $\{{\bf K}\}$ that is not commensurate with that of the final state smectic.
The system first requires a long time to disorder the initial state, and then another
long time to reorder into the smectic.  

We have checked the above results by carrying out simulations in which
we start from an initial random configuration of charges.  After roughly 
$3.5\times 10^7$ passes, we find that the system orders into the same smectic state as
in Fig.\,\ref{f25}c.  In Fig.\,\ref{f25}d, we show an intensity plot of
the real space correlations in the smectic, $C(m_1,m_2)$, obtained from the
Fourier transform of the $S({\bf k})$ of Fig.\,\ref{f25}c.  We see that charges
in the same channel ($m_2=0$) have a sharp periodic ordering.  Correlations between
channels show that the charges in neighboring channels are staggered; the peak
in charge density in one channel aligns with the minimum in charge density
of the neighboring channel, so as to form a local ordering that is more triangular
than square.  As one moves to channels further away, the correlations decrease
and the peaks in $C(m_1,m_2)$ get smeared out.

\begin{figure}
\epsfxsize=7.5truecm
\epsfbox{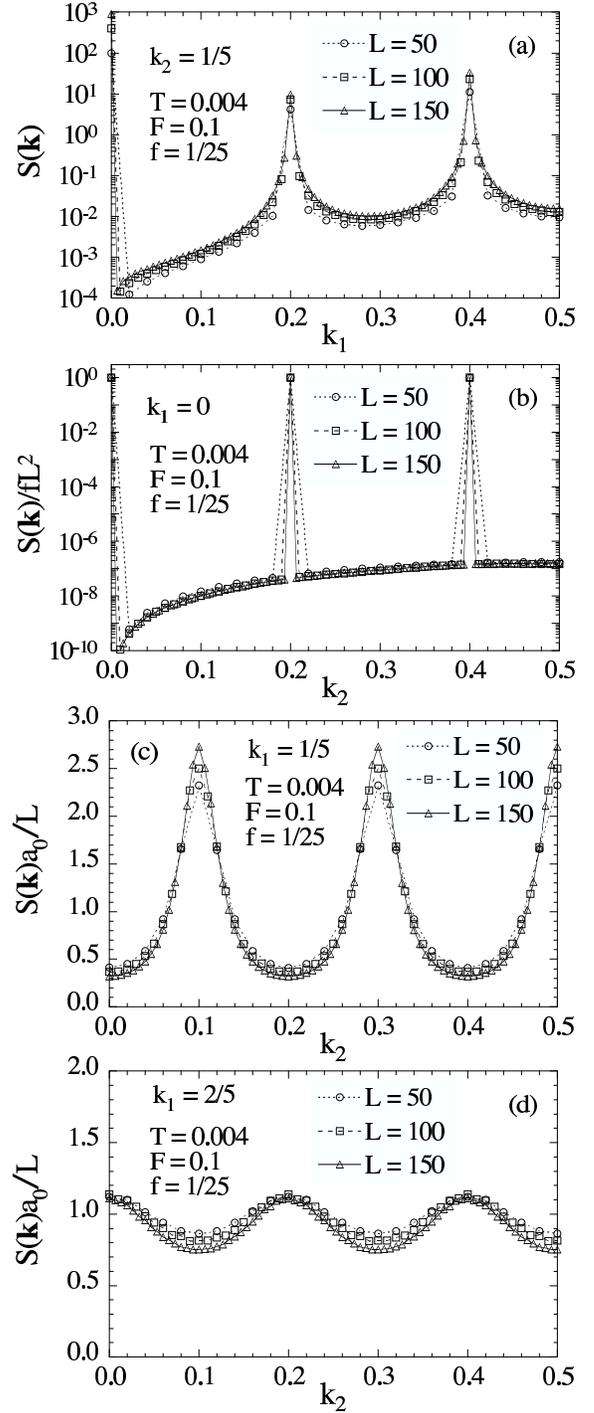}
\caption{Profiles of $S({\bf k})$ in various directions, for different system sizes $L$,
for the smectic phase at $F=0.10$, $T=0.004$ on a square grid. 
(a) $S({\bf k})$ vs. $k_1$ for fixed
$k_2=1/5$; (b) $S({\bf k})/fL^2$ vs. $k_2$ for fixed $k_1=0$;
(c) $S({\bf k})a_0/L$ vs. $k_2$ for fixed $k_1=1/5$; (d) $S({\bf k})a_0/L$
vs. $k_2$ for fixed $k_1=2/5$.  Note the logarithmic scale in (a) and (b).
}
\label{f26}
\end{figure}

We now check that the smectic phase of Fig.\,\ref{f25}c has the same
scaling behavior with system size that was found for the triangular grid.
For larger systems with $L=100-200$, it is not possible to simulate for the
very long times ($\sim 10^7$ passes) that are needed to order into the smectic
from either the ground state or a random initial state.  We therefore start with
an initial configuration that is a periodic repetition of the smectic state for
$L=50$, and simulate for only relatively short times.  For $L=100$, $150$, and $200$,
we use $10000$, $2000$, and $2000$ passes.  Our results are shown in 
Fig.\,\ref{f26}, where we plot the profiles of $S({\bf k})$ along different paths
through the first Brillouin zone.  Fig.\,\ref{f26}a shows that the speaks are
as sharply confined to the values $k_1=1/5,2/5$ as was found for the triangular
grid.  Fig.\,\ref{f26}b shows that the peaks at $k_1=0$ scale $\sim L^2$,
indicating long range smectic order.  Figs.\,\ref{f26}c,d, show that $S(k_1,k_2)$
for fixed $k_1=1/5,2/5$ scales rougly $\sim L$.  Note that the scaling collapse in
Fig.\,\ref{f26}c is not quite as nice as the corresponding Fig.\,\ref{f7}c for the
triangular grid.  Plotting the peak value $S({\bf K}_{11})$ versus $L$, similar to what
was done in Fig.\,\ref{f6}, gives $S({\bf K}_{11})\sim L^s$ with $s\approx {1.17}$.  
We believe that this value $s>1$, rather than being a signature of stronger
correlations between smectic channels, may just reflect the persistence of correlations 
introduced by our initial periodic configuration, which have not yet completely washed out
over our relatively short runs.

\begin{figure}
\epsfxsize=7.5truecm
\epsfbox{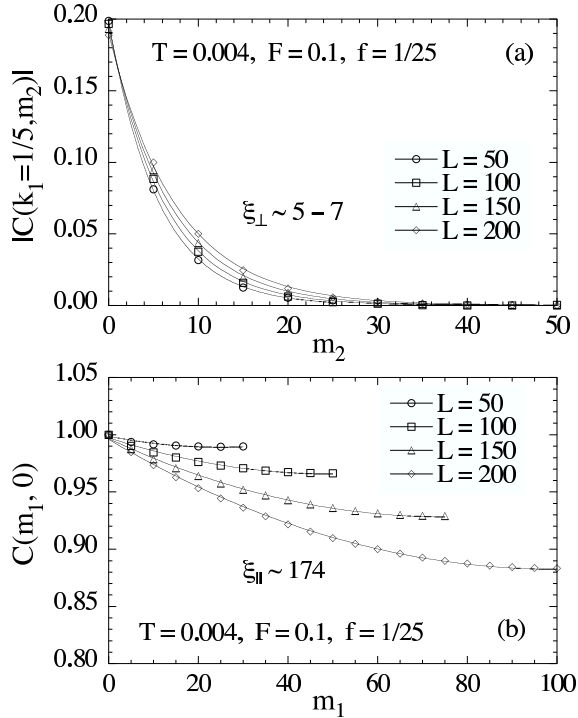}
\caption{Transverse and longitudinal correlation functions at $F=0.1$, $T=0.004$, 
for system sizes $L\times L$ on a square grid.
(a) $|C(k_1=1/5,m_2)|$ at values $m_2=5n$, $n$ integer, for different
sizes $L$; solid lines are fits to the periodic exponential of Eq.\,(\ref{ePexp}).
(b) $C(m_1, m_2=0)$ at values $m_1=5n$, $n$ integer, for different
sizes $L$; solid lines are fits to the periodic exponential of Eq.\,(\ref{ePexp2}).
}
\label{f27}
\end{figure}

In Fig.\,\ref{f27} we plot the transverse and longitudinal correlation functions,
obtained from the appropriate Fourier transform of $S({\bf k})$.  
$C(k_1=1/5,m_2)$ in Fig.\,\ref{f27}a shows exponentially decaying transverse
correlations, with a correlation length $\xi_\perp\sim 5-7$, comparable to 
that found for the triangular grid at the same parameter values.  We believe that the 
slight increase from $\xi_\perp\simeq 5$ to $7$ as $L$ increases from $50$ to $200$
reflects the correlations introduced by our initial configuration, as already
commented on in connection with Fig.\,\ref{f26}c.  Note that the peak values of
$C(k_1=1/5,m_2)$ oscillate in sign for successive values of $m_2=5n$, $n$ integer,
due to the $\pi$ phase shift from channel to channel that is apparent in the
real space correlations shown in Fig.\,\ref{f25}d; hence we have plotted
$|C(k_1=1/5,m_2)|$ in Fig.\,\ref{f26}a.  In Fig.\,\ref{f27}b we plot
the longitudinal correlation $C(m_1,m_2=0)$.  The solid lines are fits to 
a periodic exponential, and give a common value of $\xi_\|\simeq 174$ for all
sizes $L$.  Thus we have a finite correlation length, but $\xi_\|\sim L$.  
We conclude that the driven steady state for low $T$ and large $F$ on the
square grid is a smectic that is qualitatively the same as what was found for
the triangular grid.

\subsection{Low Drive}
\label{sSqLD}

\begin{figure}
\epsfxsize=7.5truecm
\epsfbox{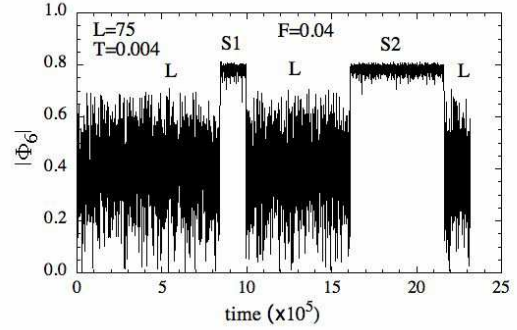}
\caption{6-fold orientational order parameter $|\Phi_6|$ vs. simulation clock time t
for $T=0.004$, $F=0.04$ and $L\times L$ system of size $L=75$.  Regions denoted
``L" are in a moving liquid state; regions denoted ``S1" and ``S2" are in a moving solid state.
Coexistence of the two states indicates that the system is at the melting transition.
}
\label{f28}
\end{figure}

\begin{figure}
\epsfxsize=7.5truecm
\epsfbox{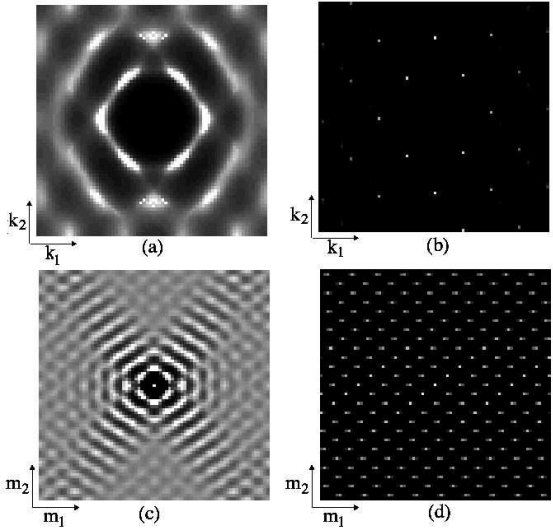}
\caption{Intensity plots of $S({\bf k})$ at $T=0.004$, $F=0.04$, for an $L\times L$
system of size $L=75$ averaged over (a) the liquid states labeled ``L" and (b) the solid states
labeled ``S2" of Fig.\,\protect\ref{f28}.  Instensity plots of the corresponding real
space correlations $C({\bf r})$ for (c) the liquid states ``L" and (d) the solid states ``S2".
The applied force ${\bf F}$ is in the horizontal direction.
}
\label{f29}
\end{figure}

We now consider behavior at low drive, simulating at $T=0.004$, $F=0.04$
for an $L\times L$ system of size $L=75$.  We will find that these parameters 
place the system right at the melting transition.
We start from an initial random configuration and run $2.5\times 10^4$ passes
to equilibrate, followed by $2.5\times 10^7$ passes to compute averages.
In Fig.\,\ref{f28} we plot the instantaneous absolute value of the 
6-fold orientational order parameter $|\Phi_6|$ versus the simulation clock
time $t$.  We see that the system makes
sharp jumps between states of lower and higher values of $|\Phi_6|$ and conclude that
these  are the coexisting
liquid and ordered phases at the first order melting transition.  In Fig.\,\ref{f29}a
we show an intensity plot of the structure function $S({\bf k})$ averaged
over only the liquid states labeled ``L" in Fig.\,\ref{f28}.  
We see a liquid like $S({\bf k})$, but with
a striking 6-fold modulation of intensity in the diffuse peaks, corresponding to
the relatively large values of $|\Phi_6|\sim 0.4$ seen in Fig.\,\ref{f28}.
In Fig.\,\ref{f29}b we show $S({\bf k})$ averaged over the ordered states labeled ``S2"
in Fig.\,\ref{f28}.  We see periodic sharp peaks suggesting a moving solid state.
$S({\bf k})$ for the states labeled ``S1" in Fig.\,\ref{f28}
is identical to that of Fig.\,\ref{f29}b, except reflected about the $k_2$ axis.
In Figs.\,\ref{f29}c,d we show intensity plots of the corresponding
real space correlations $C({\bf r})$. 


\begin{figure}
\epsfxsize=7.5truecm
\epsfbox{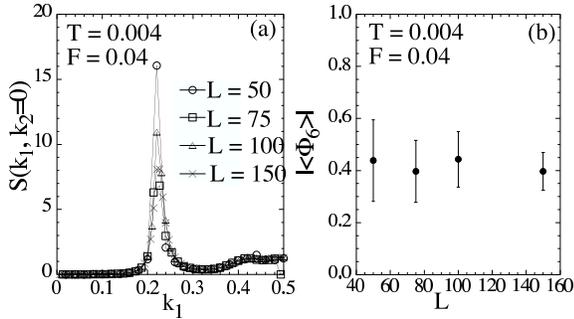}
\caption{The liquid state at $T=0.004$, $F=0.04$ for different $L\times L$ system sizes. 
(a) $S(k_1,k_2=0)$ vs. $k_1$;
(b) 6-fold orientational order parameter $|\langle\Phi_6\rangle|$ vs. $L$; bars denote the
standard deviation of the distribution of $|\Phi_6|$.
}
\label{f30}
\end{figure}

We consider first the liquid state. In Fig.\,\ref{f30}a we plot $S(k_1,k_2=0)$
versus $k_1$ for several different $L\times L$ system sizes.  Except for the
maximum of the first peak, we see essentially no finite size effect.  The height of the
first peak is different for the different $L$, however there is no systematic
variation with $L$; we believe that these differences are just statistical fluctuations.
We conclude that this state is a liquid with only short range translational order.
Figs.\,\ref{f28} and \ref{f29}a however suggest that the liquid may possess finite
orientational order.  In Fig.\,\ref{f30}b we therefore plot $|\langle\Phi_6\rangle|$
versus $L$ for the liquid state.  We see that  $|\langle\Phi_6\rangle|$ is roughly
independent of $L$, confirming that the liquid has long range hexatic orientational
order.

\begin{figure}
\epsfxsize=7.5truecm
\epsfbox{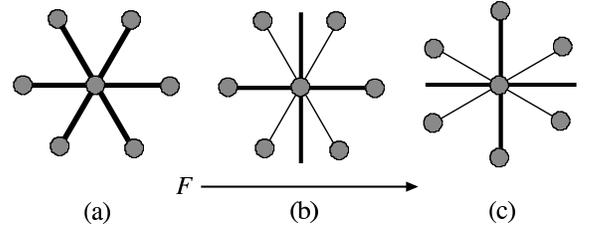}
\caption{Schematic showing how particle clusters with local 6-fold
orientational order (shaded circles connected by thin lines) may align
with the rotational symmetry of an external potential or grid (thick lines).
(a) lock in of the cluster to a 6-fold rotationally invariant triangular grid;
(b) lock in of the cluster to the horizontal axis of a 4-fold rotationally
invariant square grid; and (c) lock in to the vertical axis of a square grid.
}
\label{f31}
\end{figure}

The long ranged hexatic liquid that we find in Fig.\,\ref{f29}a is reminiscent of the
long ranged hexatic liquid found for the triangular grid at low temperatures, as shown in
Fig.\,\ref{f18}.  There are, however, some important differences.  A liquid
in a continuum will always have {\it local} 6-fold orientational order.  
However, due to the short ranged translational correlations, the phase of the local complex
orientational order parameter will vary with position.  For a normal
liquid, this causes correlations of the 6-fold orientational order parameter to decay exponentially
with distance.  According to the theory of melting in two dimensions by Halperin and Nelson,
and by Young \cite{HNY}, there may also be an algebraically ordered 
hexatic liquid between the solid and normal liquid
phases, in which correlations of the 6-fold orientational order parameter decay
algebraically.  When the system sits on an external periodic potential, however, the
local 6-fold orientational order parameter can lock onto the symmetry directions
of the external potential, which therefore serves as an ordering field for orientational 
order.  For a triangular grid, the local 6-fold order of the particles locks onto the
6-fold rotational symmetry of the grid, as illustrated in Fig.\,\ref{f31}a.  
The result is long range 6-fold orientational order, with a finite $\langle\Phi_6\rangle$.
For a square grid, the local 6-fold order of the particles may lock onto {\it either}
the vertical {\it or} the horizontal directions of the grid, as illustrated in Figs.\,\ref{f31}b,c.
For a liquid in equilibrium, both of these orientations will occur in equal numbers
on average.  Since the two orientations are related by a $\pi/2$ rotation, the
relative phase of the 6-fold orientational order parameter for the two cases is
${\rm exp}(i6\pi/2)=-1$, and adding them in equal numbers causes $\langle\Phi_6\rangle$
to vanish.  A square grid induces no 6-fold orientational order in equilibrium.
For a liquid in a driven non-equilibrium steady state, however, the direction of
the driving force ${\bf F}$ breaks the symmetry between the vertical and horizontal
directions, and can cause one to be favored over the other.  A driving force therefore
can lead to a finite $\langle \Phi_6\rangle$ and long range 6-fold orientational order
on the square grid.  From the plot of the real space correlation $C({\bf r})$
shown in Fig.\,\ref{f29}c, we see that the system locks onto the {\it vertical}
direction, as in Fig.\,\ref{f31}c.  The resulting structure function $S({\bf k})$,
shown in Fig.\,\ref{f29}a has a set of 6 peaks about the origin, which are oriented
so that one pair of the peaks align with the direction parallel to the applied driving force
${\bf F}$.  This is in contrast to case for the triangular grid, shown in Fig.\,\ref{f18},
where the peaks are oriented so that one pair of the peaks align with the direction
transverse to the applied force.

We now consider the ordered moving state, labeled ``S2" in Fig.\,\ref{f28}.
The structure function $S({\bf k})$, and the real space correlations $C({\bf r})$
are shown in Figs.\,\ref{f29}b and \ref{f29}d respectively.  Note that the
periodic peaks in $S({\bf k})$ for this ordered state do {\it not} have the
same symmetry as that of the equilibrium ground state.  The latter (see Fig.\,\ref{f25}a)
consists of a {\it square} array of Bragg peaks, while in Fig.\,\ref{f29}b the peaks are distorted
into a more triangular structure.  From either the location of the peaks in $S({\bf k})$,
or more easily from a direct inspection of the real space correlation $C({\bf r})$,
we see that this state consists of periodic channels of 
charges oriented parallel to the applied force ${\bf F}$ in the $\hat a_1=\hat x$
direction.  Within each channel the charges are ordered with 
an average separation of $8{1\over 3}$
grid spacing, while the channels themselves are separated from each other by $3$ grid spacings.
The nearest neighbors to a given charge are located in its two nearest neighboring channels, rather 
than within the same channel, reflecting a similar orientation of hexatic order
as in the liquid.
This can be compared with the smectic state at high drive, shown in Fig.\,\ref{f25}d.
This smectic has channels in which charges are separated by $5$ spaces, while the 
channels themselves are separated by $5$ spaces; the nearest neighbors to a given charge
are located within the same channel.   In contrast, the equilibrium ground state of Fig.\,\ref{f24}a
can be thought of as channels in which charges are separated by $25$ spaces, while
the channels themselves are separated by $1$ space; nearest neighbor charges lie in the
next-next-nearest neigbhoring channels.  
The structure in terms of channels, as described above, is determined by the 
strength of the correlations between the channels.  When channels are more strongly correlated,
it can be energetically favorable to have the channels spaced more closely together, with
a correspondingly larger distance between charges within a given channel; the stronger
correlations between the channels will keep charges within neighboring channels from
approaching each other too closely.  The equilibrum ground state represents the extreme
limit of long range correlations between channels.  The high drive smectic represents the
opposite limit where correlations between channels are very short ranged and it becomes
favorable to keep the spacing between channels at the same distance as the spacing of
charges within a channel.
The moving state of Fig.\,\ref{f29}d
can thus be thought of as having a structure, and presumably channel correlations,
intermediate between these two limits.

\begin{figure}
\epsfxsize=8.5truecm
\epsfbox{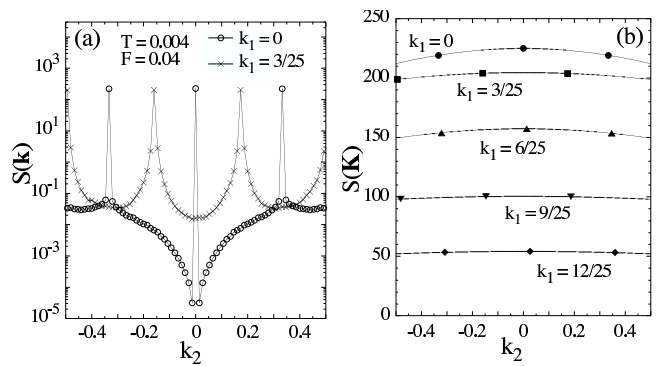}
\caption{Moving ordered phase of Fig.\,\protect\ref{f29}b, for $T=0.004$,
$F=0.04$, and $L=75$. (a) Profiles of $S(k_1,k_2)$ versus $k_2$ showing
the peaks at $k_1=0$ and $k_1=3/25$; note the logarithmic scale. 
(b) Heights of the dominant peaks in 
$S(k_1,k_2)$ versus $k_2$; the different curves represent different values of $k_1$ and
the solid lines are fits to a Gaussian.
}
\label{f32}
\end{figure}

The strong correlations between channels, as discussed above, are clearly evident
in the plots of $S({\bf k})$ and $C({\bf r})$ in Figs.\,\ref{f29}b and \ref{f29}d.
All peaks in $S({\bf k})$ appear sharp in both 
the longitudinal and transverse directions; real space correlations $C({\bf r})$
appear to extend the entire length of the system in both
longitudinal and transverse directions.  This suggests a moving solid rather than a smectic.
To investigate this further we consider in more detail the peaks in the structure
function $S({\bf k})$.  In Fig.\,\ref{f32}a we plot profiles of $S(k_1,k_2)$
versus $k_2$, showing the peaks at $k_1=0$ and $k_1=3/25$.
For $k_1=0$ the peaks appear as sharp $\delta$-function peaks upon a smooth 
background, similar to what
was seen in Fig.\,\ref{f26}b for the smectic at large drive, indicating the periodicity
of the channels in the direction transverse to the driving force ${\bf F}$.
At finite $k_1=3/25$ the peaks are much sharper
than the corresponding finite $k_1$ peaks for the smectic at large drive
in Fig.\,\ref{f26}c; in the present case the peaks drop by three orders of magnitude
from maximum to minimum (note the logarithmic scale) and have a half
width of about $\Delta k\simeq 0.007$.  Such sharp peaks
suggest the possible presence of long ranged or algebraic correlations between
the channels.

\begin{figure}
\epsfxsize=8.5truecm
\epsfbox{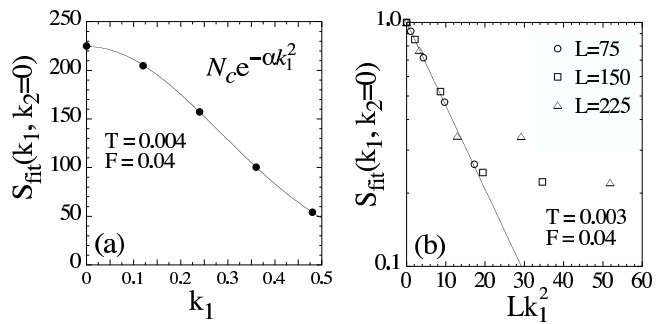}
\caption{(a) Moving ordered phase of Fig.\,\protect\ref{f29}b, for $T=0.004$,
$F=0.04$, and $L=75$. 
Values $S_{\rm fit}(k_1,k_2=0)$ of fits  from Fig.\,\protect\ref{f32}b 
vs. $k_1$. Solid line is a fit of the data to a Gaussian. (b) Moving ordered
phase for $T=0.003$, $F=0.04$, and sizes $L=75$, $150$, and $225$.
Values of $S_{\rm fit}(k_1,k_2=0)$ vs. $Lk_1^2$.  Solid line is a
fit of the small $k_1$ data to a straight line.
}
\label{f33}
\end{figure}

In Fig.\,\ref{f32}b we plot only the heights of the dominant peaks in
$S(k_1,k_2)$ versus $k_2$, for the different values of $k_1$.
We see that at fixed $k_1$, there is only a very small variation of the
peak heights with
$k_2$; however the dependence on $k_1$ is considerable.  
Fitting the points for each value of $k_1$ to a simple Gaussian
(the solid lines in Fig.\,\ref{f32}b) we plot the resulting $S_{\rm fit}(k_1,k_2=0)$
versus $k_1$ in Fig.\,\ref{f33}a, where another simple Gaussian, 
$S_{\rm fit}(k_1,0)=N_c\exp(-\alpha k_1^2)$, gives an excellent fit (the solid
line in Fig.\,\ref{f33}a).  Such a Gaussian shape for the peak heights
is consistent with a Debye-Waller-like behavior for thermal fluctuations
of a solid.  


However to investigate more precisely the nature of
translational correlations, we need to investigate the dependence of
the peak heights on system size $L$.  We have {\it not}, however, been able to do 
this at $T=0.004$; when, for larger system sizes $L$, we start the system
off in an initial disordered state, we were unable to see a similar transition
to an ordered state as was found in Fig.\,\ref{f28} for $L=75$.  We 
assume that this is either because the melting temperature becomes somewhat
lower for larger $L$ (as was seen for the triangular grid), or perhaps because
the free energy barrier between the liquid and ordered states increases with
$L$ and we have not run sufficiently long to have a thermal excitation over
this barrier.

We choose, therefore, to investigate the finite size behavior at the lower
temperature $T=0.003$, taking as an initial state an appropriate cut out of,
or periodic extension of, the ordered state we found for $T=0.004$.
For sizes $L=50$ and $L=100$, when we started the system in such an initial
state, we found that the system quickly melted to a liquid.  We believe that this
is because these values of $L$ are not commensurate with the spacing of 
$3$ grid spaces between channels required by this ordered structure.
Ordered systems of size $L=75$, $150$ and $225$, however, remained stable.
Proceeding similarly to Fig.\,\ref{f32}b, we examin the peak heights of $S(k_1,k_2)$
versus $k_2$ for the various $k_1$.  At this lower temperature $T=0.003$, the
variation with $k_2$ is even smaller than that seen in Fig.\,\ref{f32}b at $T=0.004$.
Fitting to a Gaussian, we determine the values of $S_{\rm fit}(k_1,k_2=0)$ and fit these
to a Gaussian, $S_{\rm fit}(k_1,0)=N_c\exp(-\alpha (L) k_1^2)$, where $N_c=fL^2$.
We find the surprising result that $\alpha (L)\sim L$.  To show this, we plot in
Fig.\,\ref{f33}b $S_{\rm fit}(k_1,0)/N_c$, on a log scale, versus $Lk_1^2$.  We see that
the data at small $k_1$ give an excellent collapse to a straight line.  This
exponential decrease in $S({\bf K})/N_c$ with increasing $L$ suggests that the 
observed moving solid may not persist as a stable state in the large $L$ limit.

\begin{figure}
\epsfxsize=8.5truecm
\epsfbox{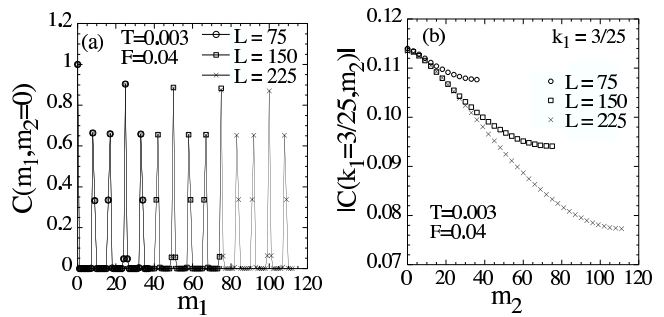}
\caption{Real space correlationf for the moving ordered phase at $T=0.003$, $F=0.04$, and $L=75$, $150$, $225$. 
(a) Longitudinal correlation $C(m_1,m_2=0)$ vs. $m_1$.
(b) Transverse correlation $|C(k_1=3/25,m_2)|$ vs. $m_2$, for $m_2=3n$, $n$ integer.
}
\label{f34}
\end{figure}

We can also examine the translational order from the perspective of the real space
correlations.  In Fig.\,\ref{f34}a we plot the longitudinal correlations, $C(m_1,m_2=0)$
versus $m_1$,
for system sizes $L=75$, $150$ and $225$.  We clearly see that there are three
charges for every $25$ grid spacings.  No finite size effect is seen.  In Fig.\,\ref{f34}b
we plot the absolute value of the complex correlation $C(k_1=3/25, m_2)$ versus $m_2$,
showing values for only every third grid spacing, $m_2=3n$, $n$ integer.
Here we find a pronounced finite size effect, with the correlation decaying to lower
values as $L$ increases.  However a periodic exponential (as used for example in Fig.\,\ref{f27}a)
does not give a particularly good fit, and we do not have enough sizes $L$ to try any 
systematic scaling fit.  While the results of Figs.\,\ref{f33}b and \ref{f34}b thus
suggest that long range solid order may not persist as $L$ increases, larger sizes
will be needed to clarify the true large $L$ behavior.

The structure that we have found in Fig.\,\ref{f29}b,d for the ordered moving
state at low drive has neither the commensurability with respect to the underlying grid of
the equilibrium ground state, nor the large drive smectic.  One can speculate
that at other values of $F$ and $T$, in this low temperature ordered region, yet
other commensurabilities may be found.  Exploring the complete phase diagram of the
driven lattice Coulomb gas on the square grid may therefore prove to be considerably 
more challenging than was for the case of the triangular grid, and we leave this
for future investigations.

\subsection{Dynamics}

We now consider  some of the dynamic properties for the driven 
Coulomb gas on the square grid.  We will not attempt a detailed 
calculation of diffusion constants, 
as we did for the triangular grid, however we will still be able to make some interesting
observations by looking at average velocity and center of mass motion.
We first consider the case of the high drive smectic, $F=0.10$, $T=0.004$, considered 
in section \ref{sSqHD}. In Table \ref{t2} we give the values for the average
center of mass velocity parallel to the driving force, $v_{{\rm ave}\,x}$, for various 
system sizes $L\times L$.  Similar to our results for the triangular grid
(see Table \ref{t1}) we find $v_{{\rm ave}\,x}\sim L$ scales proportional
to the length of the system in the direction of the applied force, in agreement with
the discussion at the end of section \ref{sLowT}.  Inspection of the center of mass
motion as a function of time clearly shows no transverse diffusion, indicating that
the smectic is transversely pinned, just as we found for the triangular grid.

\begin{table}[htdp]
\caption{Average velocity $v_{{\rm ave}\,x}$ in the high drive smectic state
for various system sizes $L\times L$ at $F=0.10$, $T=0.004$, on the square grid.}
\begin{ruledtabular}
\begin{tabular}{|c|c|c|c|}
$L$ &100 & 150 & 200 \\
$v_{{\rm ave}\,x}$ & 1082 & 1617 & 2065 \\
\end{tabular}
\end{ruledtabular}
\label{t2}
\end{table}

\begin{figure}
\epsfxsize=7.5truecm
\epsfbox{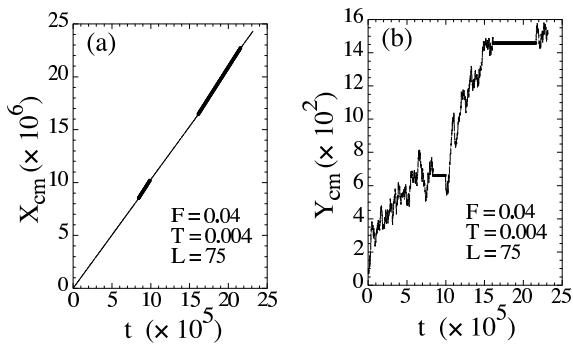}
\caption{Center of mass displacement for $F=0.04$, $T=0.004$, and system size $L=75$,
at the melting transition: (a) motion parallel to ${\bf F}$, $X_{\rm cm}$ vs. time $t$;
(b) motion transverse to ${\bf F}$, $Y_{\rm cm}$ vs $t$.  Light lines correspond to
times when the system is in the liquid state.  Heavy lines correspond to times when the
system is the ordered state (see Fig.\,\ref{f28}).
}
\label{f35}
\end{figure}

Next we consider the case of low drive, $F=0.04$, considered in section \ref{sSqLD}.
We consider first the case at melting, $T=0.004$ and $L=75$, where the system is making transitions between the liquid and a more ordered state, as shown in Fig.\,\ref{f28}.
In Fig.\,\ref{f35}a we plot the component of the instantaneous center of mass displacement
parallel to the
driving force, $X_{\rm cm}$, versus the simulation clock time $t$.  Light lines denote times
in which the system is in the liquid state, while heavy lines denote times when the system
in the ordered state (compare with Fig.\,\ref{f28}).  
That the lines in each region of time are perfectly straight indicates a
constant average velocity $v_{{\rm ave}\,x}$ in each region.  Note, however, that
the velocity in the ordered state is slightly larger than that in the liquid state: in the 
liquid, $v_{{\rm ave}\,x}=10.05$, while in the ordered state, $v_{{\rm ave}\,x}=11.50$,
or $14\%$ larger.  In Fig.\,\ref{f35}b we plot the transverse component of the
center of mass displacement, $Y_{\rm cm}$, versus the simulation clock time $t$.
In the liquid, $Y_{\rm c}$ shows the noisy fluctuations characteristic of diffusion; the
observed bias towards increasing values of $Y_{\rm cm}$ we believe is just a statistical
fluctuation.  In the ordered phase, however, $Y_{\rm cm}$ stays essentially constant
indicating that the system is transversely pinned.

We now consider the ordered phase at the lower temperature $T=0.003$.
In Table \ref{t3} we give the values of $v_{{\rm ave}\,x}$ for systems of
different sizes $L\times L$.  We consider only the values $L=75$, $150$ and $225$
that result in an ordered moving state.  In contrast to the high drive smectic (see 
Table \ref{t2}) we now find that  $v_{{\rm ave}\,x}$ is independent of $L$.
This at first seems paradoxical, since the ordered state at low drive is more strongly
correlated than the smectic at high drive.  However it may be that the incommeasurability
of the ordered state in the parallel direction (where the average spacing between
charges is $8{1\over 3}$ grid spacings) is sufficient to remove the energy barriers
responsible for the avalanche effects (see section \ref{sLowT}) that give rise to
the  $v_{{\rm ave}\,x}\sim L_1$ dependence in the high drive smectic.
Finally, an examination of the transverse displacement, similar to that of
Fig.\,\ref{f35}b, shows that the ordered state is transversely pinned.

\begin{table}[htdp]
\caption{Average velocity $v_{{\rm ave}\,x}$ in the ordered phase
at various system sizes $L\times L$
for $F=0.04$, $T=0.003$, on the square grid.}
\begin{ruledtabular}
\begin{tabular}{|c|c|c|c|}
$L$ & 75 & 150 & 225 \\
$v_{{\rm ave}\,x}$ & 36.3 & 36.3 & 36.1 \\
\end{tabular}
\end{ruledtabular}
\label{t3}
\end{table}

\section{Discussion and Conclusions}
\label{sConc}

In this work we have applied lattice gas dynamics to model the
non-equilibrium steady states of a driven diffusive system, the 2D classical
lattice Coulomb gas in a uniform applied force.  We have considered two different 
dynamic algorithms, and have found that they result in qualitatively different phase 
diagrams, contrary to naive expectations.
We have shown that the commonly used driven diffusive 
Metropolis Monte Carlo algorithm (DDMMC) results in a structurally disordered 
moving steady state over most of the phase diagram.  We have argued that this
is due to unphysical intrinsic randomness in the algorithm that remains even
as $T\to 0$.  

We have then applied continuous time Monte Carlo (CTMC) to
the driven diffusive problem and found it to produce smectic and, for the 
square grid, possibly more strongly correlated steady states at low temperatures.
We have shown (Appendix A) that CTMC is a natural discretization of continuum
Langevin dynamics.  We have argued (section \ref{sCTMC}) that in general it gives a physically 
correct dynamics when the grid sites are regarded as the minima of an external one
body potential $U({\bf r})$, and the energy barriers $U_0$ of this potential remain larger than
the energy change $\Delta E$ in hopping between neighboring minima, so that
motion is by thermal activation of one particle at a time over the potential barriers.  
It remains unclear whether or not CTMC will be qualitatively correct in the very large
drive limit, $\Delta E>U_0$, when the applied force overcomes the pinning force of the
potential, and the minima of the corresponding washboard potential, 
$U({\bf r})-{\bf F}\cdot{\bf r}$, vanish.
In such a case, for a {\it spatially uniform} system in an initially ordered state,
each charge will experience an equal net force forward from the washboard potential,
and one would expect at low temperatures that the charges would move coherently 
together.  The CTMC algorithm, which only moves a single charge at a time, 
breaks this spatial uniformity and might introduce unphysical effects.  For the
case of a system with {\it quenched randomness}, however, the random pinning
already breaks spatially uniformity, and the forces on the charges will in general
be different.  In such a case, the single particle moves of the CTMC algorithm
may not be as unphysical.
This very large drive limit for the case of random pinning
has been the subject of numerous recent theoretical \cite{Giamarchi, Balents, Scheidl,Koshelev} 
and numerical \cite{Moon, Koshelev, Faleski, Ryu, Spencer, Dominguez, Kolton, Fangohr} 
works.  

For CTMC we have shown (section \ref{sHigh}) that diverging correlation lengths as $T\to 0$
can give rise to subtle finite size effects that can be difficult to detect with
the usual finite size scaling methods applied to the peaks of the structure
function, $S({\bf K})$, and we have argued that the smectic state that we find for
finite size systems will become unstable to a liquid on sufficiently large length scales.
However, since the relevant correlation lengths diverge as $T\to 0$, the smectic
will be the stable steady state in any finite system, at sufficiently low temperature.
We have also shown that, on a square grid, long range hexatic orientational order
develops in the moving steady state liquid, and that this is a purely non-equilibrium effect.

The one component 2D lattice Coulomb gas  serves as a model for logarithmically
interacting point vortices in a 2D superconducting network, or a superconducting film
with a periodic potential.  Driven vortices in a 2D periodic potential at {\it finite}
temperature have been simulated by several others using {\it continuum} 
dynamics.  The molecular dynamic simulations of Reichhardt and 
Zim\'anyi \cite{Reichhardt} and  of Carneiro \cite{Carneiro} used square
periodic pins embedded in a flat continuum, with a number
of vortices equal to, or greater than, the number of pins.  We do not expect
that such models,
in which a sizable fraction of the vortices spend most of their time in the
flat space between the pins, will be well described by our dilute density
of charges on a discrete grid, where all charges spend most of their time
at the potential minima.

Much closer to our model is that of Marconi and Dom\'{\i}nguez \cite{Marconi1,Marconi}
who simulate the dynamics of a square array of Josephson junctions using 
resistively-shunted-junction (RSJ) dynamics applied to a 2D XY model.
They study a vortex density per unit cell of the array of $f=1/25$, the same
density as used in our present work.  The phase diagram which they
report has some qualitative similarity to our own phase diagram of Fig.\,\ref{f3}b,
with an ordered, transversely pinned, moving state at low temperatures \cite{note2}.
However, in contrast to either the smectic we find at high drives (see Fig.\,\ref{f25}c)
or the more ordered state we find at low drives (see Fig.\,\ref{f29}b), they find
a moving vortex lattice where $S({\bf k})$ has peaks at the {\it same}
reciprocal lattice vectors ${\bf K}$ as the equilibrium ground state.
From a finite size scaling analysis of $S({\bf K})$ using $L=50$, $100$, and $150$,
they conclude that their state is a vortex lattice with anisotropic algebraically decaying 
translational correlations.

To understand possible reasons for the difference between their results
and ours, we first consider the relevant parameters of their model.
For their cosine Josephson junction model, the effective one body
potential \cite{Lobb} that the array structure introduces for vortex motion has
an energy barrier $U_0\simeq 0.12$, in units where the Josephson
coupling energy is $J_0=1$.  Many of Marconi and Dom\'{i}nguez's results are in the limit
where the force $F$ (i.e. the applied current in the Josephson array model)
satisfies $F>U_0$.  This is the case where the minima in the washboard potential
parallel to the driving force have vanished, and where we have argued that
our lattice gas dynamics might not apply.  However, even for the case
$F<U_0$, the two models may be in different parameter regimes.
Our lattice gas dynamics implicitly assumes that the energy barrier
$U_0$ is larger than all other energy scales.  For the Josephson array
of Marconi and Dom\'{i}nguez, however, a direct calculation using
the XY model shows that the  energy to move a single 
vortex forward one grid space from its ground state position is $\Delta E_1\simeq 0.34$,
substantially bigger than the barrier $U_0\simeq 0.12$.   Our simulations
are therefore in the limit of a much stronger pinning potential.

In spite of these parameter differences, we can still make some
observations.  First we note that Marconi and Dom\'{i}nguez always begin
their simulations from the equilibrium ground state (or states evolved from it);
they are unable to cool the system from a liquid and find the
ordered state, hence there is no independent check that the state they find is
the true stable steady state.  Next, we note that because their simulations
use a continuum dynamics, they are unable to simulate for the very long times
that are possible using our lattice gas dynamics.  As a measure of the effective
simulation time, we can compute the total displacement $\Delta R_{\rm cm}$
of the vortex center of mass parallel to the applied force over the total time of 
the simulation.  For the Josephson array, if
$V$ is the average measured voltage drop per junction parallel to the applied
current, $I_0$ the critical current of a single junction, $R_N$ the normal
shunt resistence, $f$ the vortex density, and $\tau_J\equiv\hbar/(2eR_NI_0)$ the
time constant, then $\Delta R_{\rm cm}=(V/I_0R_N)(\Delta t/\tau_J)(N_{t}/2\pi f)$,
where $\Delta t$ is the time integration step of the simulation, and $N_{t}$
is the number of such steps.  Using Marconi and Dom\'{i}nguez's values \cite{Marconi} of
$\Delta t/\tau_J =0.1$, $f=1/25$, $N_t=10^5$, and typical values \cite{Marconi} of 
$V/I_0R_N$ from their Fig.\,5, we find for their simulations that 
$\Delta R_{\rm cm}\sim 1.2\times 10^3$ grid spacings or less.  In contrast, our simulations which lead to
Fig.\,\ref{f25}c have a total simulation time corresponding to $\Delta R_{\rm cm}\simeq
6\times 10^7$ grid spacings, more than 4 orders of magnitude larger.
To make a better comparison with Marconi and Dom\'{i}nguez,
we note that our results of Fig.\,\ref{f25}a, starting from the equilibrium
ground state, correspond to a total center of mass displacement of only
$\Delta R_{\rm cm}\simeq 4\times 10^3$
grid spacings, similar to that of Marconi and Dom\'{i}nguez.
The state we find in Fig.\,\ref{f25}a has peaks in $S({\bf k})$ at the
same ${\bf K}$ as the equilibrium ground state, moreover the anisotropies
of this state are the same as for the state found by Marconi and 
Dom\'{i}nguez; the peaks develop a finite width in the direction transverse
to the direction of motion (this feature is visible in Fig.\,\ref{f25}a),
and the heights of the peaks decrease as ${\bf k}$ varies in the direction
of motion.  In our case the variation in peak heights is only a $20\%$
reduction from largest to smallest, whereas for Marconi and Dom\'{i}nguez
it is a larger $75\%$, nevertheless the behavior is qualitatively similar.
It thus may be that the simulations of Marconi and Dom\'{i}nguez have 
not run long enough to observe the true long time steady state of the system.

Finally, we comment on one additional issue that is related to our ability,
using lattice gas dynamics, to simulate to much longer times that can be
achieved with continuum methods.
It is interesting to note in Fig.\,\ref{f23} that the longitudinal diffusion constant
$D_{xx}$ in the liquid approaches its long time limit on a much longer time scale than does
the transverse diffusion constant $D_{yy}$.  In recent continuum Langevin
simulations \cite{Kolton2} of driven vortices in a disordered 2D superconductor, 
similar diffusion in the vortex liquid phase was 
computed.  Although it was observed that the transverse diffusion constant saturated to
a finite value at long times, the longitudinal diffusion constant was found not to saturate,
but rather to grow proportional to $t$.  
Rather than reflecting super-diffusive behavior
in the longitudinal direction \cite{Kolton2}, this result might simply be
a failure to simulate to long enough times to see the longitudinal diffusion
constant saturate.

\section*{Acknowledgements}

We wish to thank D.~Dom\'{\i}nguez, M.~C.~Marchetti and A.~A.~Middleton
for helpful discussions.  This work was supported by DOE grant DE-FG02-89ER14017
and NSF grant PHY-9987413.

\section*{Appendix A}

In this appendix we demonstrate that the transition rates of Eq.\,(\ref{eWialpha})
correctly describe diffusive Langevin motion in the limit that the energy
change in one move satisfies $\Delta U \ll T$.  Our derivation follows one
given earlier \cite{Teitel} for a single degree of freedom, and extends it to the
case of many degrees of freedom.

The Langevin equation of motion for diffusively moving particles
in a uniform driving force ${\bf F}$ can be written as,
\begin{equation}
{\partial r_{i\alpha} \over\partial t} = - D{\partial U\over\partial r_{i\alpha}} +\eta_{i\alpha}
\enspace,
\label{eLangevin}
\end{equation}
where $r_{i\alpha}$ is the $\alpha$ component of the position of particle $i$, 
\begin{equation}
U[\{{\bf r}_i\}]\equiv {\cal H}[\{{\bf r}_i\}]-{\bf F}\cdot\sum_i {\bf r}_i
\enspace,
\label{eU}
\end{equation}
with ${\cal H}$ the Hamiltonian giving the internal interactions between
the particles, and $\eta_{i\alpha}$ is the $\alpha$ component of the
thermally fluctuating force acting on particle $i$.  In order that
the system reaches equilibrium in the absence of the force ${\bf F}$, the
thermal force is taken to have correlations,  
\begin{eqnarray}
\langle\eta_{i\alpha}(t)\rangle &=&0\\
\langle\eta_{i\alpha}(t)\eta_{j\beta}(t^\prime)\rangle&=&2DT
\delta_{ij}\delta_{\alpha\beta}\delta(t-t^\prime)\enspace.
\label{eeta}
\end{eqnarray}

The corresponding Fokker-Planck equation that describes the probability
$P(\{{\bf r}_i\})$ for the system to be at coordinates $\{{\bf r}_i\}$
is then given by,
\begin{equation}
{\partial P\over\partial t}=D\sum_{i\alpha}\left[{\partial\over\partial r_{i\alpha}}
\left(P{\partial U\over\partial r_{i\alpha}}\right)+T
{\partial^2P\over\partial r_{i\alpha}^2}\right]\enspace.
\label{eFP}
\end{equation}

Next we symmetrize the Fokker-Planck equation by making the transformation,
\begin{equation}
\psi(\{{\bf r}_i\})\equiv {\rm e}^{U[\{{\bf r}_i\}]/2T}P(\{{\bf r}_i\})\enspace.
\label{epsi}
\end{equation}
Substituting the above into the Fokker-Planck equation (\ref{eFP}) gives
the imaginary time Schr\"odinger equation,
\begin{equation}
{\partial \psi\over\partial t}=DT\sum_{i\alpha}\left[{\partial^2\psi\over\partial r_{i\alpha}^2}
-V_{i\alpha}\psi\right]\enspace,
\label{eSchrodinger}
\end{equation}
where,
\begin{equation}
V_{i\alpha}=\left[\left({1\over 2T}{\partial U\over\partial r_{i\alpha}}\right)^2
-{1\over 2T}{\partial^2U\over\partial r_{i\alpha}^2}\right]\enspace.
\label{eV}
\end{equation}

If we now discretize the coordinates, so that the ${\bf r}_{i}$ are confined
to the sites of a periodic lattice, the natural way to discretize 
Eq.\,(\ref{eSchrodinger}) is to replace the
second derivative of $\psi$ with its lattice equivalent,
\begin{equation}
{\partial \psi\over\partial t}=DT\sum_{i}\left[\Delta^2_i-V_i\right]\psi\enspace,
\label{eSDiscrete}
\end{equation}
%
%
where $\Delta^2_i$ is the discrete Laplacian for the lattice with respect to
coordinate ${\bf r}_i$, and $V_i$ the appropriate discretization of $\sum_\alpha V_{i\alpha}$, 
as will be explained below.  For a lattice with nearest neighbors given
by the vectors $\{\pm \hat a_\mu\}$,
the discrete Laplacian acting on a scalar function $f({\bf r})$ is defined by,
\begin{equation}
\Delta^2f({\bf r}) \equiv c\sum_\mu \left[ f({\bf r}+\hat a_\mu)-2f({\bf r}) +f({\bf r}-\hat a_\mu\right)]\enspace,
\label{eDL}
\end{equation}
with $c$ an appropriate geometrical constant to give the correct continuum limit.

If we denote the state of the system with particles at positions $\{{\bf r}_i\}$ as $s$,
then we can write the above Eq.\,(\ref{eSDiscrete}) in a matrix form,
\begin{equation}
{\partial\psi_s\over\partial t}=\sum_{s^\prime}\tilde M_{ss^\prime}\psi_{s^\prime}\enspace,
\label{eM}
\end{equation}
where the matrix $\tilde{\bf M}$ has elements,
\begin{eqnarray}
\tilde M_{ss}&=&-DT\left[zc+V_i\right]\label{eMt00}\\
\tilde M_{ss^\prime}&=&cDT,\quad{\rm when}\> s^\prime 
=\{{\bf r}_i\pm\hat a_\mu,{\bf r}_j\}\label{eMt}\\
\tilde M_{ss^\prime}&=&0,\quad{\rm otherwise}\label{eMt2}\enspace,
\end{eqnarray}
where $z$ is the number of nearest neighbor sites.
By the notation in Eqs.\,(\ref{eMt}) and (\ref{eMt2}) we mean that
the only non-zero off-diagonal elements of $\tilde{\bf M}$ are those
connecting states $s$ and $s^\prime$ in which only a single particle at ${\bf r}_i$ has moved to a 
nearest neighbor position ${\bf r}_i\pm\hat\alpha_\mu$, and all other particles have
remained unchanged.
This is our first result: the natural discretization of continuum Langevin dynamics to a 
lattice gas dynamics involves single particle moves only.

To see what are the correct single particle hopping rates for our lattice gas dynamics,
as well as to see what is the correct discretized form for the $V_i$ of Eq.\,(\ref{eSDiscrete}),
consider now the Master Equation for our lattice gas dynamics.  If
$s=\{{\bf r}_i\}$ is the state of the system, then the probability $P_s$ to be in state
$s$ is determined by,
\begin{equation}
{\partial P_s\over\partial t} = \sum_{s^\prime}\left[W_{ss^\prime}P_{s^\prime}-
W_{s^\prime s}P_s\right]\equiv \sum_{s^\prime}M_{ss^\prime}P_{s^\prime}\enspace,
\label{eMaster}
\end{equation}
where $W_{s^\prime s}$ is the rate to make a transition from state $s$ to state
$s^\prime$.  We therefore have,
\begin{eqnarray}
M_{ss}&=&-\sum_{s^\prime}W_{s^\prime s}\label{eMdiag0}\\
M_{ss^\prime}&=&W_{ss^\prime}, \quad s\ne s^\prime \enspace.
\label{eM2}
\end{eqnarray}

We  next apply the same transformation as in Eq.\,(\ref{epsi}), to get,
\begin{equation}
{\partial\psi_s\over\partial t} = \sum_{ss^\prime}{\rm e}^{U_s/2T}M_{ss^\prime}
{\rm e}^{-U_{s^\prime}/2T}\psi_{s^\prime}\enspace.
\label{epsi2}
\end{equation}
Comparing with Eq.\,(\ref{eM}) we get,
\begin{equation}
\tilde M_{ss^\prime}={\rm e}^{(U_s-U_{s^\prime})/2T}M_{ss^\prime}\enspace,
\label{eM3a}
\end{equation}
and from Eqs.\,(\ref{eMt}) and (\ref{eM2}) we then get
for the {\it off-diagonal} elements of $\tilde{\bf M}$,
\begin{equation}
\tilde M_{ss^\prime}={\rm e}^{(U_s-U_{s^\prime})/2T}W_{ss^\prime} =cDT\enspace,
\label{eM3b}
\end{equation}
when the state $s^\prime$ differs from the state $s$ by only a single particle that has moved
to a nearest neighbor site, i.e. ${\bf r}_i\to {\bf r}_i\pm\hat a_\mu$, with all other ${\bf r}_j$
kept constant; all other off-diagonal terms vanish.  The above result then 
determines the transition rates that are needed for the discrete Master Equation to model the
continuum Langevin equation,
\begin{equation}
W_{ss^\prime}=cDT{\rm e}^{-(U_s-U_{s^\prime})/2T}\enspace.
\label{eW}
\end{equation}
Thus we arrive at the rates of Eq.\,(\ref{eWialpha}) that define our CTMC algorithm.

Note that the rates of Eq.\,(\ref{eW}) satisfy a local detailed balance,
\begin{equation}
{W_{ss^\prime}\over W_{s^\prime s}} ={\rm e}^{-(U_s-U_{s^\prime})/T}\enspace.
\label{eDBalance}
\end{equation}

Having determined the rates $W_{ss^\prime}$, we can now determine the 
diagonal part of $\tilde{\bf M}$ and hence the $V_i$ of Eq.\,(\ref{eSDiscrete}).
Defining $\Delta U_{i\pm\mu}$ as the change in $U$ when a single particle
moves ${\bf r}_i\to{\bf r}_i\pm\hat a_\mu$, i.e.,
\begin{equation}
\Delta U_{i\pm\mu}\equiv U[\{{\bf r}_i\pm\hat a_\mu,{\bf r}_j\})]
-U[\{{\bf r}_i,{\bf r}_j\}]\enspace,
\label{eDU}
\end{equation}
then Eqs.\,(\ref{eMdiag0}) and (\ref{eM3a}) give for the diagonal elements of $\tilde M_{ss}$,
\begin{eqnarray}
\tilde M_{ss}&=&M_{ss}=-\sum_{s^\prime}W_{s^\prime s}\\
                   &=&-cDT\sum_{s^\prime}{\rm e}^{-(U_{s^\prime}-U_{s})/2T}\\
                   &=&-cDT\sum_{i\mu}\left[ {\rm e}^{-\Delta U_{i+\mu}/2T}
                   +{\rm e}^{-\Delta U_{i-\mu}/2T}\right]\enspace.
\label{eMdiag}
\end{eqnarray}
%
If one now expands Eq.\,(\ref{eMdiag}) to order $(\Delta U/2T)^2$, and then uses 
Eq.\,(\ref{eDL}) that
$c\sum _\mu(\Delta U_{i+\mu}+\Delta U_{i-\mu})=\Delta_i^2 U$, and compares
to Eq.\,(\ref{eMt00}), one concludes that, 
\begin{equation}
V_i=-{\Delta_i^2 U\over 2T}+\sum_\mu\left[
{c\over 2}\left({\Delta U_{i+\mu}\over 2T}\right)^2
+{c\over 2}\left({\Delta U_{i-\mu}\over 2T}\right)^2\right]\enspace.
\label{eVi}
\end{equation}
This is just the natural symmetric discretization of $V_i=\sum_\alpha V_{i\alpha}$
with $V_{i\alpha}$ given by Eq.\,(\ref{eV}).

We have thus shown that the CTMC dynamics, with rates as in Eq.\,(\ref{eWialpha}),
is the natural discretization of overdamped Langevin dynamic in the continuum, and
that CTMC becomes a very good approximation for the continuum dynamics in the
limit that the energy changes for single particle moves, $\Delta U_{i\alpha}$ of
Eq.\,(\ref{eDU}), satisfy $\Delta U_{i\alpha}\ll 2T$.



\begin{thebibliography}{99}
\bibitem{note} In the very large drive limit $F\gg U_0$, when the work done by
the force dominates over the change in interaction energy, it may be possible
to choose $W_0$ dependent on $F$ (and possibly varying with grid direction
$\hat a_\mu$) in such a way as to again match onto the continuum solution.
This remains for further investigation.
\end{thebibliography}
\end{document}